\documentclass[fleqn,usenatbib]{mnras}

\usepackage{newtxtext,newtxmath}
\usepackage[T1]{fontenc}

\DeclareRobustCommand{\VAN}[3]{#2}
\let\VANthebibliography\thebibliography
\def\thebibliography{\DeclareRobustCommand{\VAN}[3]{##3}\VANthebibliography}

\usepackage{graphicx}	% Including figure files
\usepackage{amsmath}	% Advanced maths commands

\usepackage{booktabs}  % create tables with elegant horizontal lines.
\usepackage[authoryear]{natbib}
\usepackage{chngcntr}  % numbering to be assigned as the appendix format: Figure A.1
\usepackage{hyperref}
\hypersetup{
    citecolor=blue,
    colorlinks=true,
    linkcolor=blue,
    filecolor=magenta,      
    urlcolor=blue,
    pdftitle={Overleaf Example},
    pdfpagemode=FullScreen,
    }
\usepackage{CJKutf8}

\defcitealias{rs75}{RS}

\newcommand{\sect}[1]{Section~\ref{#1}}
\newcommand{\fig}[1]{Fig.~\ref{#1}}
\newcommand{\tab}[1]{Table~\ref{#1}}
\newcommand{\equ}[1]{Eq.~(\ref{#1})}
\newcommand{\apdx}[1]{Appendix~\ref{#1}}

\title[Emission beats in PSR J1514$-$4834]{The Thousand-Pulsar-Array programme on MeerKAT XVII: Discovery of beating radio emission variability in PSR J1514$-$4834}

\author[J.~A. Hsu et al.]{
J.~A. Hsu (\begin{CJK*}{UTF8}{bsmi}許睿安\end{CJK*}),$^{1}$\thanks{E-mail: jui-an.hsu@manchester.ac.uk}
P. Weltevrede,$^{1}$
G. Wright,$^{1}$
M.~J. Keith,$^{1}$
L.~S.~Oswald$^{3}$,
X. Song$^{2}$ and
\newauthor
H. Wang (\begin{CJK*}{UTF8}{bsmi}汪昊月\end{CJK*})$^{1}$
\\
$^{1}$Jodrell Bank Centre for Astrophysics, Department of Physics and Astronomy, University of Manchester, Manchester M13 9PL, UK \\
$^{2}$ASTRON, The Netherlands Institute for Radio Astronomy, Oude Hoogeveensedijk 4, 7991 PD, Dwingeloo, the Netherlands \\
$^{3}$School of Physics \& Astronomy, University of Southampton, Southampton SO17 1BJ, UK \\
}

\date{Accepted XXX. Received YYY; in original form ZZZ}

\pubyear{\the\year{}}

\begin{document}
\label{firstpage}
\pagerange{\pageref{firstpage}--\pageref{lastpage}}
\maketitle

\begin{abstract}
We present a comprehensive analysis of the complex subpulse modulation patterns in PSR~J1514$-$4834 (B1510$-$48) using L-band data from the Thousand-Pulsar-Array (TPA) programme, complemented with further MeerKAT UHF-band data.
We demonstrate that periodic drifting subpulses and rapid amplitude modulation with a period of about $2$ pulse periods coexist.
It is established that these two periodic emission patterns interfere in the form of a beat system, giving rise to multiple spectral features. We develop a new methodology which confirms the expected correlations in the complex phase of the beat features in a 2D Fourier Transform of the data. Therefore, a relatively simple beat system can explain the complex single-pulse behaviour of this pulsar.
The simultaneous coexistence of multiple subpulse modulation periodicities is rare in the population and points to poorly understood intricate dynamics within pulsar magnetospheres. A comparison with PSR~B0943$+$10 is made, for which the coexistence of multiple periodicities has been argued to be a natural consequence of a rotating carousel. However, our observations of PSR~J1514$-$4834 require a different explanation, involving time-delayed interactions between separate regions of the magnetosphere. The developed phase correlation methodology opens up the route for underlying beat systems in the modulation patterns of more pulsars.
\end{abstract}


%%%%%%%%%%%%%%%%%%%%%%%%%%%%%%%%%%%%%%%%%%%%%%%%%%

%%%%%%%%%%%%%%%%% BODY OF PAPER %%%%%%%%%%%%%%%%%%

\section{Introduction}
\label{sec:intro}

The Thousand-Pulsar-Array (TPA) programme \citep{jkk+20} is a part of the MeerTIME Large Survey Project \citep{bja+20} on the MeerKAT telescope.
Exploiting the strength of the great sensitivity of MeerKAT, the TPA has observed more than 1200 non-recycled pulsars at declinations below $\sim+20^\circ$ and regularly monitors $\approx500$ of these \citep{swk+21}.
This extensive sample has yielded rich datasets on both integrated \cite[e.g.][]{okp+21, pkj+21, pkj+23, jkk+23, jmk+24, bwk+24, kjk+24, kjp+24} and single-pulse properties.
The latter, assembled by \cite{sws+23}, includes the largest survey of subpulse\footnote{i.e. identifiable emission substructures within single pulses \citep{dc68}} modulation to date.

In this subpulse modulation survey 1198 pulsars are analysed, many exhibiting the phenomenon of drifting subpulses, as would be expected from previous surveys \cite[e.g.][]{ran86, wes06, wse07, bmm+19}. Subpulse drift occurs when, over time, distinct subpulses shift in pulse longitude within the pulse profile window \citep{ssp+70} and can be seen as periodic phase modulation.
In addition, a number of pulsars show periodic variations in pulse intensity from pulse to pulse, without accompanying phase drift, and this is known as amplitude modulation \citep[e.g.][]{wes06, bmm+16, sws+23}.
It is debated whether they are from the same physical mechanism or not \citep{bmm+19, bmm20, sws+23}.

\begin{figure}
\centering
\includegraphics[width=\hsize]{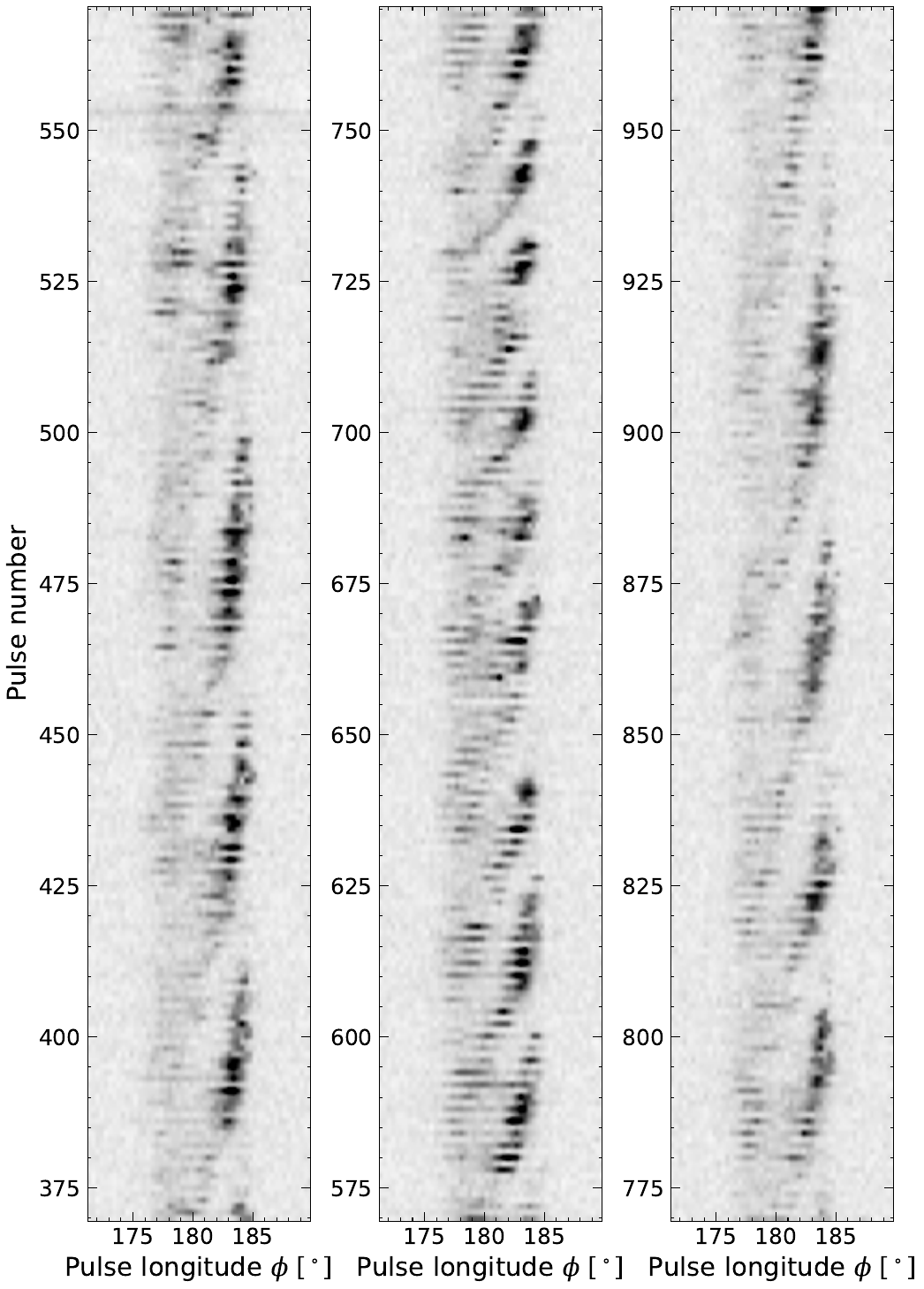}
  \caption{A section of the pulse stack of PSR~J1514$-$4834 using the L-band data. A rapid bright-weak amplitude modulation with a period $\approx 2P$ (flicker) can be seen on top of the inclined drift bands.
  In the middle panel, which we refer to as the ``event'', the drifting subpulses repeat faster ($P_3\approx24$) compared to the ordinary behaviour ($P_3\approx40$).
  The flux density is clipped at 50 per cent of the maximum to enhance the visibility of the flicker.}
  \label{fig:J1514-4834_TPA_pulsestack}
\end{figure}

PSR~J1514$-$4834 (B1510$-$48) stands out as one of the most interesting sources identified in the work of \cite{sws+23}, showing both drifting subpulses and amplitude modulation {\em simultaneously} (see \fig{fig:J1514-4834_TPA_pulsestack}).
Superimposed on the drift bands, the amplitude modulation is periodic and rapid. This takes the form of a bright-weak modulation with a period of $\approx2P$, i.e. alternating weak/bright pulses, where $P\approx 0.45~\rm s$ is the rotation period of the pulsar \citep{PSRCAT, lbs+20}.
The much slower drifting subpulses signal was also identified by \cite{bmm20} and \cite{lds+24}, but the rapid bright-weak modulation was not.
This pulsar is located near the middle of the bulk of normal pulsars in a $P$-$\dot{P}$ diagram, resulting in an inferred surface magnetic field strength $B_s\approx0.66\times10^{12}~\rm G$, as well as the inferred characteristic age $\tau_c\approx7.8\times10^6$ years and spin-down energy loss rate $\dot E\approx3.9\times10^{32}~\rm erg/s$.

In the survey of \cite{sws+23}, several pulsars show evidence of exhibiting multiple periodicities in their subpulse modulation. These mostly reflect drift mode changes, an unusual but well-known phenomenon whereby the drifting subpulse patterns switch between two or more modes over time \citep[e.g.][]{bmh+85, now91, ggk+04, svw+22}. However, the simultaneous appearance of multiple periodicities in the emission pattern of PSR~J1514$-$4834 is extremely rare in the pulsar population and will be shown here to be distinctly different from drift mode changes. It thus poses a challenge to existing theoretical models.

An early and often cited explanation for drifting subpulses is the rotating carousel model proposed by \cite{rs75} (henceforth the \citetalias{rs75} model).
In this model, a group of localised sparks is formed in the vacuum polar gap above the magnetic poles of pulsars.
The sparks generate charges which are accelerated and are responsible for the observed coherent radio emission at higher altitudes in the magnetosphere.
The electric potential close to the surface creates discharges via sparks.
These sparks, along with the sub-beams anchored to them, circulate around the magnetic axis relative to the neutron star surface, resulting in the appearance of drifting subpulses.
These take the form of diagonal drift bands in pulse stacks (see e.g. \fig{fig:J1514-4834_TPA_pulsestack}).

The drift bands are characterised by two periods, $P_2$ and $P_3$, corresponding to the horizontal (pulse longitude) and vertical (pulse number) interval between adjacent drift bands, respectively.
In the framework of the rotating carousel model, $P_3$ is determined by the circulation time of the carousel $P_4$ and the number of sub-beams $n_{\rm sb}$. Ignoring the important possibility of aliasing, which will be discussed in \sect{sec:J1514-2DFS}, $P_3$ can be taken as $P_4/n_{\rm sb}$, and therefore independent of observing frequency.
In contrast, $P_2$ depends on the spacing of the sub-beams which can be expected to vary with observing frequency together with the width of the profile. See e.g. \cite{es03} for a discussion of the frequency and pulse longitude dependence of $P_2$.

Although PSR~J1514$-$4834 is unusual in possessing amplitude modulation in addition to its drift patterns, it is not unique. 
The well-known pulsar PSR~B0943$+$10 \citep{dr99,dr01,ad01,gs03}, is comparable to PSR~J1514$-$4834 in both characteristic age and spin-down energy loss rate, and is also reported to have both these features\footnote{This refers to the pulsar's B mode. It also has a Q mode with largely featureless fluctuation spectra.}. 
PSR~B0943$+$10 has a symmetrical pair of spectral features at both sides of the main ($P_3\approx 2$) drift feature. These “sidebands” are the beat frequencies of the drifting subpulses with a slow modulation.
The authors model these features as a rotating carousel consisting of 20 sub-beams of fixed but unequal intensities.
The slow circulation time ($P_4\approx 37$) then drives the slow amplitude modulation in addition to the drifting subpulses.

Sidebands were also described for PSRs~B0834$+$06 \citep{ad05} and B1857$-$26 \citep{mr08}, while an additional slow spectral feature (without sidebands) was linked to a carousel circulation time for PSRs~B1237$+$25 \citep{md14} and J2022$+$5154 \citep{cww+24}.
The inferred number of sub-beams in the carousels for these pulsars ranges from 8 to 20 sub-beams.

Periodic amplitude modulation as seen for example in PSRs~B1133$+$16 \citep{hr07} and J1819$+$1305 \citep{rw08} has been attributed to periodic nulling, whereby a pulsar suddenly ceases its regular emission for one or many stellar rotations \citep[see, e.g.,][]{rit76, ran86, gjk12}.
Nulling is often observed as a stochastic process, although several pulsars are known to show a periodicity in the occurrence of nulls \citep[e.g.,][]{hr07, hr09, rw08, gjw14}  and can therefore be regarded as an extreme instance of amplitude modulation \citep{bmm17}. Whether periodic or not, a sparsely filled rotating carousel may be responsible, with nulls arising from a line of sight intersecting a region without active sub-beams. These have been dubbed "pseudo" nulls \citep{rwr05} as opposed to "true" nulls involving cessation of the entire emission process.

When conducting a full and comprehensive analysis of the modulations present in any pulsar, fluctuation spectra are useful and powerful tools. 
These take two principal forms.
Firstly, the longitude-resolved fluctuation spectrum \cite[LRFS,][]{LRFS} is based on one-dimensional Fourier Transforms along constant pulse longitude columns (vertical direction) in the pulse stack, and reveals $P_3$ as a function of pulse longitude.
Secondly, two-dimensional fluctuation spectra \citep[2DFS,][]{es02, wes06, wse07, sws+23}, which is based on the two-dimensional Fourier Transform of the pulse stack, provide information about both $P_2$ and $P_3$. \apdx{apdx:2DFT} provides a more detailed mathematical description.

In this paper, we will argue that the observed complex emission patterns of PSR~J1514$-$4834 result from the beat between two separate modulation patterns.
An important expectation in such scenario is that complex phases of spectral features in the 2DFS, which we refer to as Fourier phases, are correlated between different spectral features, and a new methodology is developed to meaningfully compare the Fourier phases in the context of the beat system model.
For drifting subpulses, these Fourier phases have been explored in the form of "subpulse phase tracks" \citep{es02} as the complex phases of the LRFS, to characterise the average drift band shape.
It also has been used to, for example, find delays in the phase-locked modulation between the main pulse (MP) and interpulse (IP) in PSRs~B1702$-$19 \citep{wws07} and B1055$-$52 \citep{wwj12} by comparing the subpulse phase of the MP and IP.

This paper is structured as follows: \sect{sec:obs} describes the details of the observations and data processing. In \sect{sec:J1514}, we show the complex emission patterns of PSR~J1514$-$4834 in both the time and Fourier domains and introduce the beat system model.
In \sect{sec:J1514-phase}, we introduce and use the Fourier phase correlation technique to confirm the beat system interpretation.
In \sect{sec:discussion}, we discuss the potential physical origins of the beat system, and point out a possible way to interpret the observed modulation pattern as beats between two interacting carousels.
Finally, we summarise our findings in \sect{sec:conclusion}.

\section{Observation and data processing}
\label{sec:obs}
We observed PSR~J1514$-$4834 using the MeerKAT telescope at both the L-band (856 to 1712 MHz) and UHF-band (544 to 1088 MHz). The L-band data were obtained from the Thousand-Pulsar-Array (TPA) programme, while the UHF-band data were obtained from a separate proposal.

The work of \cite{sws+23} also exploited the L-band data from the TPA programme, but only a single observation consisting of 1,040 pulses from the ``census dataset'' with the full available array (in this case 59 out of a total of 64 antennas).
In this paper, we analysed all available L-band data from September 2019 to September 2024 with a total of 15,924 pulses, including the census dataset and regular monthly observations from the TPA programme using a $\approx32$-dish sub-array.
In addition, PSR~J1514$-$4834 was observed for 80 minutes using the MeerKAT UHF-band (on January 18, 2024) with the full array (in this case 58 available antennas), obtaining 10,549 pulses in a single long observation.

All data were recorded in full polarisation and in ``search mode'' by the Pulsar Timing User Supplied Equipment (PTUSE) backend \citep{bja+20} and sampled every 120.5 $\mu$s.
The bandwidth of the L- and UHF-bands are 856 and 544 MHz, split into 1024 and 4096 frequency channels, respectively.
These data were processed off-line in the MeerTIME single-pulse pipeline (Keith et al., in preparation), where the data are folded at the pulsar period using {\sc dspsr} \citep{DSPSR} to produce 1024 pulse phase bins within a pulsar period.
Depending on the observation, $\approx20$-30 per cent of the frequency channels in the L-band data, and 12 per cent in the UHF-band data, were discarded to mitigate for radio frequency interference (RFI). The same set of frequency channels were discarded for every single pulse within a given observation.
After RFI mitigation, the pulses are averaged over the full bandwidth and single pulse archives are generated.
To quantify the subpulse modulation properties, a further subpulse modulation pipeline as described in \cite{sws+23} is applied to produce the science-ready pulse stacks for subsequent analysis (see Sections~\ref{sec:J1514} and \ref{sec:J1514-phase}). This analysis uses the {\sc psrsalsa} software package\footnote{\url{https://github.com/weltevrede/psrsalsa}} \citep{psrsalsa}.

\section{The subpulse modulation of PSR~J1514$\,$--$\,$4834}
\label{sec:J1514}

\subsection{Drifting subpulses, flickers and their beats}
\label{sec:J1514-ps}

\fig{fig:J1514-4834_TPA_pulsestack} shows three consecutive sections of the pulse stack of PSR~J1514$-$4834 using the L-band data.
The drifting subpulses are clearly visible as diagonal bands of emission. In general, the drift bands repeat every $P_3\approx40$. However, in the middle panel (between pulses 570 to 769), the drift bands are significantly more shallow with a $P_3$ value of $\approx24$, which we refer to as the ``event'' in all the following text.
The significance of this rare event is further discussed in \sect{sec:J1514-fast}.
It is noticeable that the drift bands are curved such that they are steeper in the later part of the profile (the trailing and strongest pulse profile component as shown in \fig{fig:J1514-profile}).

In addition to the drifting subpulses, a rapid amplitude modulation such that every bright pulse is followed by a weak pulse is evident throughout the observation.
We will refer to this fast bright-weak modulation as ``flickering''. The Fourier-based census of subpulse modulation in the TPA data \citep{sws+23} revealed the existence of the periodicities associated with drifting subpulses and amplitude modulation in many pulsars.
From this census, it is clear that most pulsars with drifting subpulses only have one periodicity. 
Where multiple periodicities are detected, it is presumed that in most cases these are a consequence of successive drift mode changes \citep{sws+23}. 
Here, by contrast, we demonstrate that PSR~J1514$-$4834 exhibits both periodicities simultaneously, something which is much rarer in the pulsar population.

However, as a first step, we need to confirm that the flickering does not arise from a misidentification of the rotation period. If the pulsar has an interpulse, and is folded at half the true rotational period of the star, the MP and IP will be observed alternately at the same pulse longitude in successive pulses. This will give the impression of flickering at a modulation period of exactly two pulses. Such case was identified in the TPA data for PSR~J1618$-$4723 by \cite{sws+23}. However, for PSR~J1514$-$4834 this can be firmly ruled out, as the pulse profiles (and pulse stacks) constructed of the even- and odd-numbered pulses (not shown) are identical. Therefore the flickering represents an intrinsic phenomenon for PSR~J1514$-$4834.

\begin{figure}
\centering
\includegraphics[width=0.9\hsize]{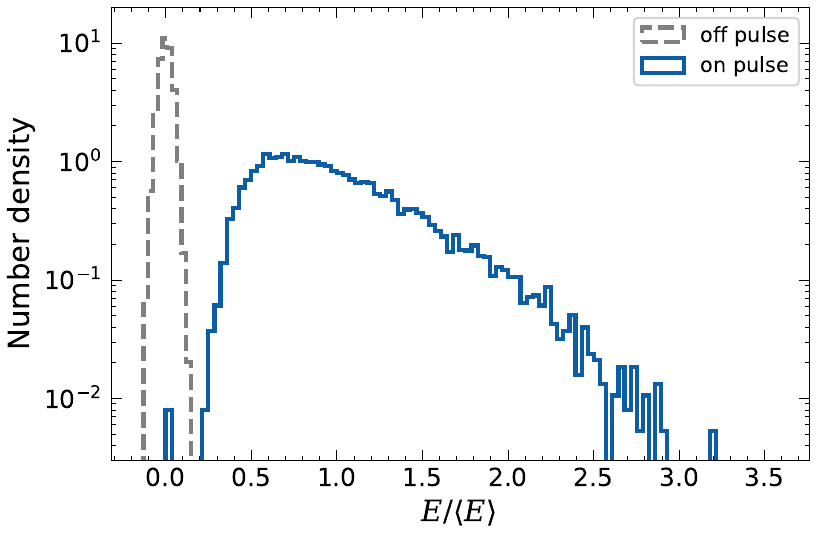}
  \caption{The pulse-energy distribution of PSR~J1514$-$4834 in the UHF-band (blue) and the off-pulse distribution (grey).
  Here, the energies are normalised to the average pulse energy $\langle E \rangle$.}
  \label{fig:J1514-energy}
\end{figure}

The observed flickering, i.e. the fast bright-weak modulation, is also very different from periodic nulling \citep[see, e.g.,][]{rit76, ran86, gjk12}.
This can be confirmed in the pulse-energy distribution obtained from the much longer UHF-band data as presented in \fig{fig:J1514-energy}.
For pulsars showing nulling, a bimodal pulse-energy distribution is expected: distinct populations of nulls and pulses where the former should be consistent with zero flux density.
However, here the pulse-energy distribution of PSR~J1514$-$4834 (blue) is almost entirely unimodal and shows a significant offset from the Gaussian-like noise distribution (grey) centred at zero. Such a continuous unimodal distribution suggests that the flicker modulation is gradual, rather than switching abruptly between distinct ``on'' and ``off'' states as expected for nulling.
Yet, apart from the persistent flickers, three nulls (blue bin centred at zero) are seen out of a total of 10,549 pulses recorded in the UHF-band. Their flux densities are consistent with the noise distribution in \fig{fig:J1514-energy} and confirmed by visual inspection of these pulses, while the remaining pulses are confidently detected\footnote{Only the UHF-band data has a energy distribution clearly separated from the noise, and as a consequence, no nulling is detected in the L-band data.}. 
Therefore, the nulling fraction of PSR~J1514$-$4834 is ultra low: $\sim0.03$ per cent. This is as low as the lowest (non-zero) nulling fraction known (PSR~J1932$+$1059; \citealt{whh+20}).
The ultra low nulling fraction of PSR~J1514$-$4834 is consistent with the upper limit in \cite{lds+24}.
No evidence is found to suggest that the occurrence of these three nulls is related to the intensity of the flicker, or the $1/P_3$ frequency of the drifting subpulses.

\subsection{The rare fast $P_3$ event}
\label{sec:J1514-fast}

Before exploring the interplay between the drifting subpulses and the flickers within the framework of the beat system model, we first investigate the extent to which the short section of data shown in \fig{fig:J1514-4834_TPA_pulsestack} characterised by a shorter $P_3\approx24$ stands out from the rest of the data.
Despite $P_3$ of the drifting subpulses being variable over time, this short $P_3$ sustained for several drift bands has been observed only in a single short stretch of data ($\approx200$ pulses) out of all L-band data (15,924 pulses) and not at all in the UHF-band data (10,549 pulses). 
Although a drift mode change would be a natural explanation for a distinct $P_3$, with only one known event in combination with the transition appearing to be more gradual, it remains unclear if this rare event should be considered to be a mode change.

\begin{figure}
\centering
\includegraphics[width=0.9\hsize]{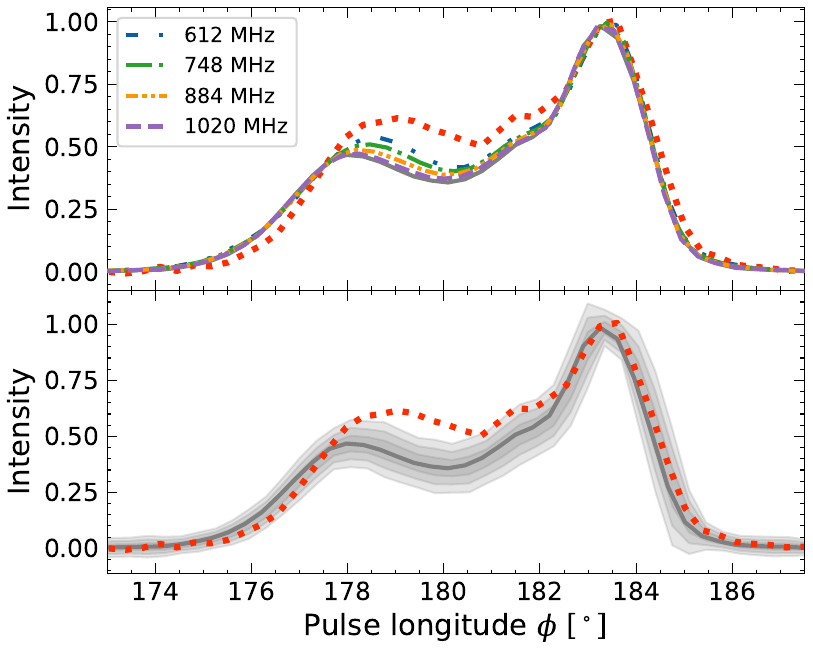}
  \caption{Pulse profiles for PSR~J1514$-$4834 normalised and aligned by their peak. 
  The red dotted lines in both panels represent the pulse profile of the rare event with $P_3\approx24$ (middle panel of \fig{fig:J1514-4834_TPA_pulsestack}).
  The L-band pulse profiles are shown as grey solid lines in both panels.
  In the upper panel, the 544 MHz bandwidth of the UHF-band is split into 4 sub-bands, labelled by their centre frequencies.
  In the lower panel, the grey shaded areas represent the distributions of pulse profile variations in all L-band observations after splitting the data into segments of 200 pulses (the length of the event). The three shades of grey correspond to $1$, $2$, and $3\sigma$.}
  \label{fig:J1514-profile}
\end{figure}

In addition to the different $P_3$ values, this event shows a distinct average pulse profile from that of the rest of the data (\fig{fig:J1514-profile}).
For better comparison, they are normalised and aligned with the peak intensity of the trailing component.
The intensity and pulse longitude of the peak are inferred by fitting the average pulse profiles with three Gaussian components.
These three components account for a relatively subtle bridge between the leading and trailing components.
For the UHF-band data (upper panel), the pulse profiles are calculated for four 136~MHz wide sub-bands separately.
The higher the observation frequency, the lower the relative intensity of the leading component.
The pulse profiles at L-band (grey line in the upper panel) and UHF-band are similar.
In contrast, the event (red dotted line in the upper panel of \fig{fig:J1514-profile}) is more distinct, with relatively strong emission at the leading part of the profile.
This difference is much bigger than can be expected from random pulse-to-pulse variability. This follows from the lower panel of \fig{fig:J1514-profile}, which compares the profile shape of the event (red dotted line) at L-band with the distribution of profiles obtained from the rest of the L-band data after dividing them into segments with the same length as the event.
The shaded contours mark the $1$, $2$, and $3\sigma$ levels of the obtained pulse profile distribution.

\begin{figure}
    \centering
     \includegraphics[width=\linewidth]{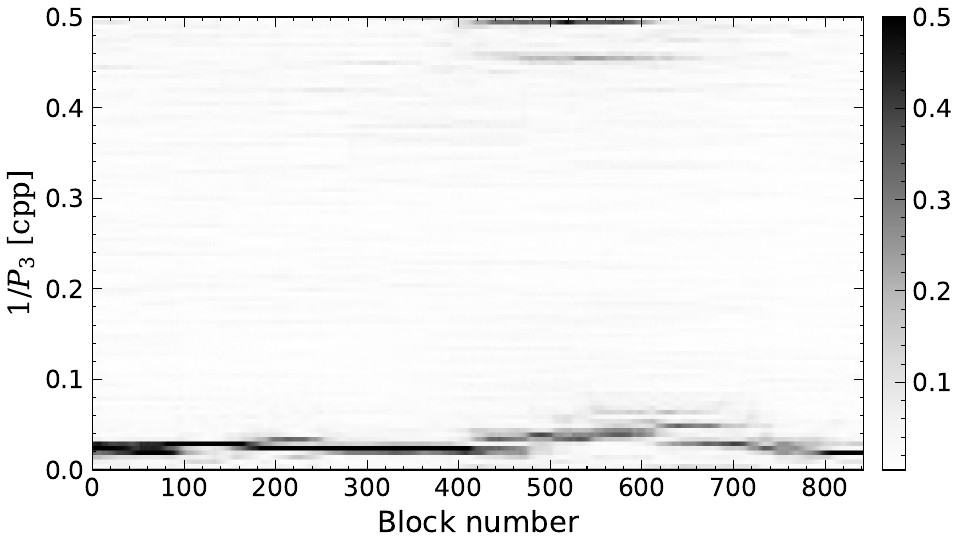}
      \caption{An S2DFS showing the $1/P_3$ evolution in cycles per period (cpp) based on one of the L-band observations in which the event occurred. Here the window width (FFT length) is 200 pulses. The spectral power is indicated by the colour bar.
      The spectral power is normalised and clipped at 50 per cent of the maximum power, allowing the weaker flicker feature to be better visible.}
      \label{fig:J1514-fast-s2DFS}
\end{figure}

Not only is the $P_3$ value of the drifting subpulses and the overall pulse profile distinctly different during the event, but the flickering also becomes more noticeable in \fig{fig:J1514-4834_TPA_pulsestack}.
A sliding two-dimensional fluctuation spectrum \citep[S2DFS,][]{ssw09} is used to quantify the strength and frequency of the flicker and drift spectral features (\fig{fig:J1514-fast-s2DFS}).
Note the power in at the DC frequency, i.e. $1/P_3 = 0$ cycles per period (cpp), has been suppressed here, as well as in all subsequent fluctuation spectra in this paper, unless stated otherwise.
An S2DFS is obtained by choosing a short pulse number window with a length equal to the FFT length and then sliding this window across the entire pulse stack.
At each window position, a 2DFS is calculated.
The power of each spectrum is then integrated in the $1/P_2$ direction to produce a $1/P_3$ distribution.
Consequently, an S2DFS provides insights into the temporal changes of $1/P_3$.

In \fig{fig:J1514-fast-s2DFS}, the drift (low-frequency) feature is strong and persistent.
The feature with $1/P_3\approx0.5~\rm cpp$, the flicker, is only strong between block numbers 400 and 600. This coincides with the time $1/P_3$ frequency of the drift feature increases, i.e. during the event.
Subsequently, the flicker fades and the $1/P_3$ frequency of the drift feature gradually returns to its normal value.
To quantify this temporal relationship, the cross-correlation between the intensity of the flicker and the $1/P_3$ frequency of the drifting subpulses was calculated.
It was found that the flicker intensity reaches its maximum preceding the peak of the $1/P_3$ frequency of the drift feature by approximately $60P$.

\begin{figure}
    \centering
     \includegraphics[width=\linewidth]{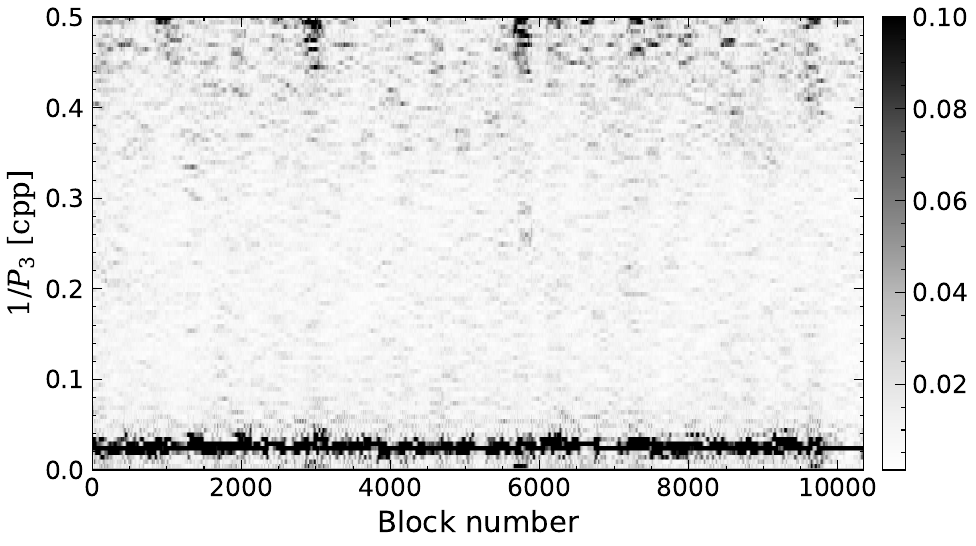}
      \caption{The S2DFS for the UHF-band data. The spectral power is clipped at 10 per cent of the maximum power, allowing the weaker flicker feature to be visible.}
      \label{fig:J1514-UHF-s2DFS}
\end{figure}

\fig{fig:J1514-UHF-s2DFS} shows the S2DFS of the much longer UHF-band data.
Both the $1/P_3$ frequency of the drift and flicker features are variable, indicating the inherent instability of the flicker and drift periodicities. 
Nevertheless, the $1/P_3$ frequency of the drift feature varies and remains within the range of the typical $1/P_3$ value of $0.025~\rm cpp$, much lower than the rare event shown in \fig{fig:J1514-fast-s2DFS}.
The ratio of spectral power of the flicker and drift features is variable and increases, for example, around block numbers 3000 and 5800.
However, no significant correlation could be established between this ratio and the $1/P_3$ frequency of the drift feature.

\subsection{Confirmation of the predicted beat features}
\label{sec:J1514-2DFS}

\begin{figure*}
    \centering
    \includegraphics[width=0.95\textwidth]{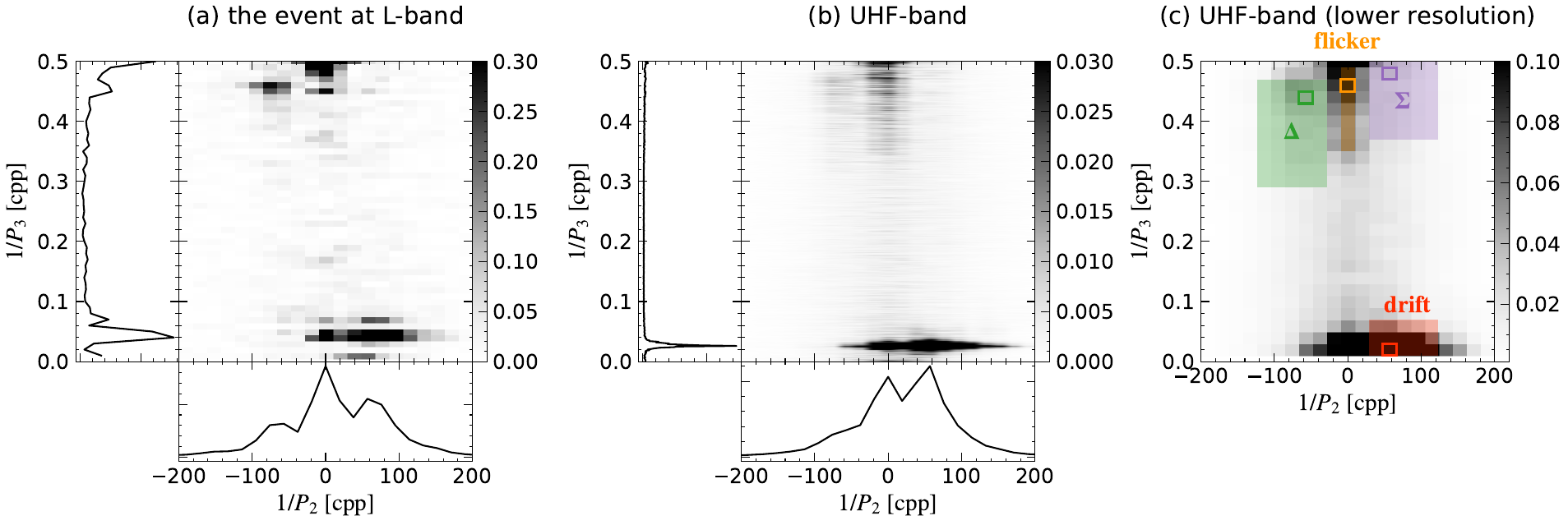}
    \caption{Power 2DFS of PSR~J1514$-$4834 based on (a) the event at L-band as described in \sect{sec:J1514-fast} and (b) the UHF-band data. Panel (c) is the same as (b), but uses a shorter FFT length. In panel (c), the drift, flicker, and beat features $\Delta$ (difference) and $\Sigma$ (sum) are indicated by shaded areas in different colours. The shaded areas and highlighted pixels (outlined boxes) are used for the Fourier phase correlation analysis (see \sect{sec:J1514-phase} and \fig{fig:J1514-phase-scatter}).
    Each spectrum is normalised separately by setting the maximum power to one.
    To highlight the weaker features, the brighter features are clipped by applying a threshold of 30 per cent, 3 per cent, and 10 per cent of the maximum spectral power in panels (a), (b) and (c) respectively.
    In panels (a) and (b), the horizontally and vertically integrated power in these two 2DFS are shown in the left and bottom side panels.
    The FFT length is 100, 1000 and 50 in panels (a), (b) and (c) respectively.}
    \label{fig:J1514-4834_2dfs}
\end{figure*}

The expected beat features are confirmed by computing the 2DFS\footnote{Following \cite{wes06,wse07}, all the 2DFS in this paper are horizontally flipped to ensure that a positive $1/P_2$ can be associated with positive drift (drift towards later pulse longitudes).} of PSR~J1514$-$4834 during the event discussed in \sect{sec:J1514-fast}.
This is shown in panel (a) of \fig{fig:J1514-4834_2dfs}.
Three spectral features are identified in panel (a), and their centroid positions are summarised in \tab{tab:measure}.
The strongest spectral feature corresponds to the drifting subpulse feature at $1/P_3\approx0.04~\rm cpp$ ($P_3\approx24$). The feature is significantly offset to positive $1/P_2$, consistent with the drift towards later pulse longitudes as seen in \fig{fig:J1514-4834_TPA_pulsestack}.
The feature at $1/P_3\approx0.49~\rm cpp$ centred at $1/P_2\approx0$ corresponds to the flickering, which does not have any significant phase drift.
A similar feature appears in Fig. 3 in \cite{lds+24}, although they do not discuss or comment on this.
In addition to these two spectral features associated with the identified periodicities in the pulse stack, there is an extra spectral feature with a negative $1/P_2\approx-75~\rm cpp$ and $1/P_3\approx0.45~\rm cpp$.
So the apparent drift direction of this feature is opposite to that of the drifting subpulses.
This feature is also visible in \fig{fig:J1514-fast-s2DFS} at the same time the flicker feature is strong.
This extra feature can be naturally attributed to a beat of the other two periodicities, and does not require additional physical complexity such as bi-drifting \cite[e.g.][]{clm+05, psrsalsa}.

\begin{table}
\caption{$1/P_3$ and $1/P_2$ measurements of the centroid locations of the three spectral features identified in panel (a) of \fig{fig:J1514-4834_2dfs} (the rare event at L-band). This is followed by the predicted location of the beat frequencies of the subpulse drift and flicker. $\Delta$ and $\Sigma$ correspond to the difference and summed features respectively. For the latter, alias is taken into account (see main text).}
\label{tab:measure}
\centering
\begin{tabular}{lcccc}
\toprule
    & $1/P_3$ & $P_3$ & $1/P_2$ & $P_2$ \\
    & [cpp] & & [cpp] & [$^\circ$] \\
\midrule
drifting subpulse       &  $0.042\substack{+0.009\\-0.005}  $  &  $24\substack{+3\\-4}$       &  $51\substack{+21\\-10}$   &  $7\pm2$       \\ [0.7em]
flicker     &  $0.49\substack{+0.01\\-0.05}$    &  $2.03\substack{+0.20\\-0.05}$  &  $0\pm10$                  &  -             \\ [0.7em]
beat        &  $0.454\substack{+0.004\\-0.033}$ &  $2.20\substack{+0.18\\-0.02}$  &  $-75\substack{+10\\-14}$  &  $-4.8\pm0.7$  \\ [0.7em]
predicted $\Delta$             &  $0.45\substack{+0.01\\-0.05}$   &  $2.23\substack{+0.28\\-0.05}$  &  $-51\substack{+10\\-23}$  &  $-7\pm2$      \\ [0.7em]
predicted $\Sigma$ (alias)     &  $0.47\substack{+0.05\\-0.01}$   &  $2.14\substack{+0.05\\-0.21}$  &  $-51\substack{+10\\-23}$  &  $-7\pm2$      \\
\bottomrule
\end{tabular}
\end{table}

If the drift and flicker periodicities are beating, then two extra spectral features are expected.
One with the sum of the frequencies (which we refer to as $\Sigma$), and one with the difference ($\Delta$).
This applies to both the $1/P_2$ and $1/P_3$ frequency, and therefore the position of the $\Sigma$ and $\Delta$ features can be predicted (see \tab{tab:measure}).
Indeed, the extra feature identified in panel (a) of \fig{fig:J1514-4834_2dfs} (labelled ``beat'' in \tab{tab:measure}) coincides with the predicted $\Delta$ feature.
For PSR~J1514$-$4834 the $\Sigma$ feature is not visible as a separate feature in panel (a) because the summed $1/P_3$ frequency exceeds the Nyquist frequency ($\approx0.532>0.5~\rm cpp$). This means the apparent frequency of $\Sigma$ would be aliased and appear in the observed 2DFS at $1/P_3\approx0.468~\rm cpp$ and $1/P_2\approx-51~\rm cpp$ in \fig{fig:J1514-4834_2dfs}. 
The predicted location of the $\Sigma$ feature as given in \tab{tab:measure} corresponds to the aliased position.
Given the spectral width of the features, the $\Delta$ and $\Sigma$ features are predicted to significantly overlap, and the single observed beat feature is formed.
The conclusion that the $\Delta$ and $\Sigma$ features are expected to be observed as a single merged feature is independent of whether the observed drift and flicker frequencies are aliased or not.

Although clearest then, the beating of drifting subpulses and flicker is not only detectable during the event. The 2DFS of the long UHF-band data is shown in panel (b) of \fig{fig:J1514-4834_2dfs}.
As described in \sect{sec:J1514-fast}, the profile evolution of PSR~J1514$-$4834 is subtle with observing frequency, and the drift patterns at UHF-band (not shown) are very similar to those at L-band with $P_3\approx40$.
As a consequence, the 2DFS of the UHF-band data is qualitatively similar to panel (a), although the flickering, and hence the beat feature, are relatively weak. The much longer observation also means that the flicker feature as well as the beat features are considerably smeared out in the $1/P_3$ (vertical) direction, indicating that the flicker frequency is inherently less stable compared to the drift frequency.
In panel (c) it is highlighted that some of the power of the broad $\Sigma$ feature is visible at positive $1/P_2$, although most is aliased and superimposed on the $\Delta$ feature on the other side.
Further strong evidence for a beat being responsible for the complex nature of the observed spectra follows from the Fourier phase correlation technique presented in the next section.

\section{Fourier phase correlation}
\label{sec:J1514-phase}

The power spectra of PSR~J1514$-$4834 are complex (see \sect{sec:J1514-2DFS} and \fig{fig:J1514-4834_2dfs}), with an additional beat feature originating from a flicker modulating the drifting subpulses.
Apart from during the short fast $P_3$ event, both flicker and beat features are significantly smeared out, and the two predicted beat features cannot be detected.
Without the observation of the rare event, establishing that a beat is responsible for the complexities in the observed 2DFS would not be possible using power spectra alone.
Here we will show that the Fourier phase information (see \apdx{apdx:2DFT} for a mathematical definition), which is discarded in a power spectrum, provides the valuable additional information.

The Fourier phases of a single 2DFS, together with the 2DFS power spectrum, fully describe the pulse stack.
These Fourier phases are in some ways analogous to subpulse phase tracks (see \apdx{apdx:2DFT}), which are the Fourier phases from an LRFS. For example, the $1/P_2$ frequency of a spectral feature in a 2DFS can also be inferred from the slope of the subpulse track associated with a specific $1/P_3$ frequency bin. The complex phases in the 2DFS (Fourier phases) indicate the vertical offset of the subpulse phase tracks (see also \fig{fig:2DFT-2dsin-conv}). 
As described in for example \cite{wws07, wwj12}, the actual values of these phases are not of physical interest, as they depend on the phase of the modulation cycle at which the first single pulse in the pulse stack happened to be recorded. 
However, differences in phase values are meaningful, and allow, for example, time delays in the modulation pattern in different profile components to be quantified \citep[e.g.][]{wws07, wwj12}.
To determine if a spectral feature is a beat, we will use the fact that the beat frequency, as well as the Fourier phase of a beat feature, is the sum or difference of the two periodicities that are beating (see \apdx{apdx:beat}).
Here we will illustrate this new technique by applying it to the UHF-band data and more L-band data (not during the event) of PSR~J1514$-$4834 where the beating is more difficult to identify.

\begin{figure*}
    \centering
    \includegraphics[width=0.95\linewidth]{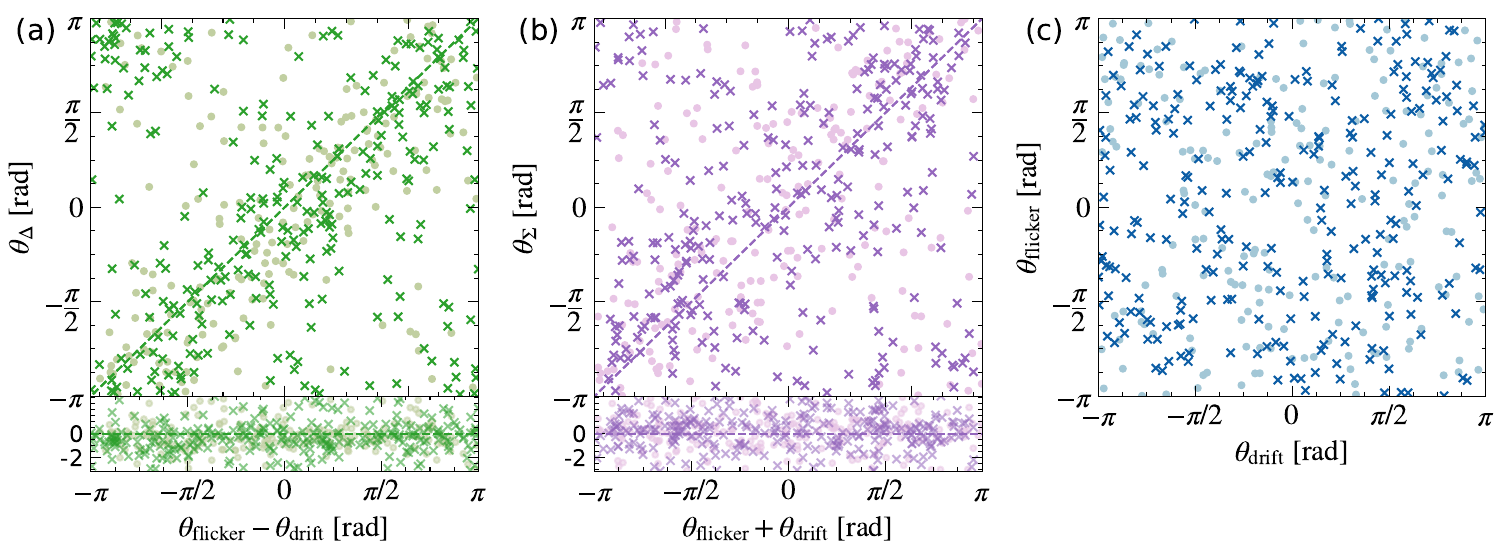}
    \caption{The scatter plots of the Fourier phase measurements from the highlighted pixels in panel (c) of \fig{fig:J1514-4834_2dfs}.
    The dots and crosses represent the UHF- and L-band data, respectively.
    Panels (a) and (b) compare the predicted Fourier phases of the beat features ($\theta_{\rm flicker}-\theta_{\rm drift}$ and $\theta_{\rm flicker}+\theta_{\rm drift}$) with the measured ones ($\theta_\Delta$ and $\theta_\Sigma$).
    The Fourier phases of the drift and flicker features, assumed to be uncorrelated, are plotted in panel (c).
    The dashed lines correspond to a one-to-one relation.
    Below each scatter plot, the residuals relative to the dashed lines are displayed.
    As predicted, the correlations in the first two panels are significant, and absent in the third (see main text).}
    \label{fig:J1514-phase-scatter}
\end{figure*}

The procedure for calculating the Fourier phase correlation between spectral features is as follows (see also \apdx{apdx:WSRT} for an application to synthetic data): for a sufficiently long observation, the entire pulse stack can be segmented into several blocks with a length equal to the FFT length which determines the spectral resolution. 
Each block produces a complex Fourier spectrum, including both the power and phase information. 
Although it is common practice to average power spectra (shown in \fig{fig:J1514-4834_2dfs}), averaging phase spectra is not meaningful.
Furthermore, subpulse modulation frequencies are generally intrinsically unstable, adding complexity to the phase analysis.
Hence, we instead compare the Fourier phase of the presumed beat feature to the predicted phase based on a measurement of the phases of drift and flicker features in individual short blocks of data.
To do this, we first select two pixels in the 2DFS representing the drift and flicker features, which are shown in panel (c) of \fig{fig:J1514-4834_2dfs} by the red and orange outlined boxes, respectively.
The corresponding positions of the beat feature pixels $\Delta$ and $\Sigma$ follow automatically from this pixel choice as the sum and difference of the frequencies of drift and flicker features, as indicated by the green ($\Delta$) and purple ($\Sigma$) outlined boxes.
The phase relationship between these four pixels can then be investigated for correlations as presented below.

The Fourier phases of these four pixels, i.e. $\theta_{\rm drift}$, $\theta_{\rm flicker}$, $\theta_\Delta$, and $\theta_\Sigma$, are compared in the scatter plots in \fig{fig:J1514-phase-scatter}.
For each segment of data containing 50 pulses for which a fluctuation spectrum is computed (therefore the FFT length is also 50 pulses), a single data point is generated in these scatter plots.
Data points from the UHF- and L-band data obtained from separate observations are shown in the same figure.
Both are based on the same FFT length and pixel selection as indicated in panel (c) of \fig{fig:J1514-4834_2dfs}.
In the first two plots of \fig{fig:J1514-phase-scatter}, the vertical axes are the measured Fourier phases of the beat features $\theta_\Delta$ and $\theta_\Sigma$. 
The horizontal axes are the corresponding predicted phases based on the measured Fourier phases of the drift and flicker features ($\theta_{\rm flicker} - \theta_{\rm drift}$ and $\theta_{\rm flicker} + \theta_{\rm drift}$).
If the spectral features $\Delta$ and $\Sigma$ were unrelated to the drifting subpulse and flicker features, their Fourier phases would be uncorrelated with the predicted phases.
However, in panels (a) and (b), 
the measured phases of both beat features follow a one-to-one relationship with the prediction (which we will show to be significant). This is true for both the UHF- and L-band data.
This firmly establishes that the spectral features identified as beat features are indeed closely related to drifting subpulse and flicker features, and confirms the prediction of the beat system interpretation.
The residual plots below the scatter plots show that, as expected, the residuals relative to the model’s predictions cluster around zero.

For completeness, the Fourier phases of the drifting subpulse and flicker features are compared in panel (c). They are uncorrelated and therefore randomly scattered points are observed.
As a result, the drift and flicker features contain independent information.
More pairs of spectral features are compared in \apdx{apdx:corr_other}.
From this it follows that $\theta_\Delta$ and $\theta_\Sigma$ are only correlated with the specific combinations of $\theta_{\rm flicker}$ and $\theta_{\rm drift}$ shown in \fig{fig:J1514-phase-scatter}.

The significance of the correlations observed in \fig{fig:J1514-phase-scatter} can be quantified with the circular correlation coefficient $r$ (see \apdx{apdx:r_fl} for its definition).
This accounts for the cyclic nature of the phases which are compared.
Given the similarity between the UHF- and L-band data shown in \fig{fig:J1514-phase-scatter}, they are used collectively in the correlation analysis.
Both panels (a) and (b) show a positive correlation with $r=0.25$ and $r=0.13$.
In contrast, the uncorrelated data in panel (c) has a much smaller $r=-0.0005$.
To statistically describe the significance of the measured correlation, statistical hypothesis tests are performed and the calculated $p$-values for the null hypothesis that posits no correlation (one-sided test\footnote{In this case, the $p$-value represents the probability of observing a correlation as strong as, or stronger than, the one found in the sample, assuming that there is actually no correlation in the population.}) are $3\times10^{-54}$, $6\times10^{-29}$, and $0.40$ for panels (a)-(c), respectively.
Consequently, the probability that the beat features are uncorrelated with the drift and flicker features is extremely small, while, as expected the phases of the drifting subpulses and flickers are uncorrelated. 
Given the $\Sigma$ pixel (purple outlined box in panel (c) of \fig{fig:J1514-4834_2dfs}) has lower spectral power compared to the $\Delta$ pixel (green outlined box), it is no surprise that the correlation in panel (a) of \fig{fig:J1514-phase-scatter} is stronger than in panel (b).

The Fourier phase correlation analysis presented in \fig{fig:J1514-phase-scatter} is based on a specific pixel combination (the outlined boxes in panel (c)). 
However, because of intrinsic variability in the periodicities, as well as other effects, spectral features are broadened and the expected correlation exists in more pixel combinations.
Therefore, a range of pixel combinations could be used. This is done for the shaded regions in panel (c) of \fig{fig:J1514-4834_2dfs} as discussed in \apdx{apdx:corr_other}.
These results are consistent with \fig{fig:J1514-phase-scatter}, confirming that a beat system is in general active in PSR~J1514$-$4834, both during the rare fast $P_3$ event (see \sect{sec:J1514-2DFS}) and at other times. 
Moreover, it is shown in \apdx{apdx:corr_other} that also the presence of the aliased $\Sigma$ feature which overlaps with the $\Delta$ feature can be established with the Fourier phase correlation methodology.

\section{Discussion}
\label{sec:discussion}

In recent times, time-domain studies of radio pulsar emission have proved crucial for advancing our understanding of the dynamics within the pulsar magnetosphere.
Yet questions have remained unanswered about the physical origins of and relationships between different time-scales of emission variability.
Detailed studies of unusual pulsar behaviour, including this work, are indispensable as they have the potential to test existing theories and provide crucial constraints on possible models.
Therefore, in this section, we will explore the physical interpretation of the complex emission of PSR~J1514$-$4834 based on the observations and analysis results presented in previous sections.

After discussing how this pulsar is placed in a wider population, we show that at least two distinct periodicities are required to explain the observed complexity. 
Note an additional independent process may be involved to account for the sporadic nulls, which are distinct from the flickers for this pulsar.
However, 
we do not discuss it further in the following discussion.
It is argued that the coexistence of drifting subpulses and flickers is incompatible with the traditional circulation model.
As an alternative, we confirm that these phenomena can be successfully encompassed by a feedback model.

\subsection{PSR~J1514$-$4834 and the pulsar population}
\label{sec:population}

In the survey of 1198 pulsars conducted by \cite{sws+23}, 35 per cent had detectable drifting subpulses, including PSR~J1514$-$4834. The $P_3$ of the drifting subpulses of PSR~J1514$-$4834 (also in the rare fast $P_3$ event) falls within the large range of $P_3$ values of other pulsars with similar $P$ and $\dot P$.
Periodic amplitude modulation, without any evidence for drift, is less common in the pulsar population.
\cite{sws+23} reported 181 pulsars (15 per cent) with periodic amplitude modulation, with little dependence on $P$ and $\dot P$.
The phenomenon of flicker is rare within the pulsar population: among the 181 pulsars with amplitude modulation identified by \cite{sws+23}, only 21 pulsars (9 per cent) show amplitude modulations with $P_3 \le 2.5$.
Note that the flicker feature in the 2DFS of PSR~J1514$-$4834 was not classified as amplitude modulation in \cite{sws+23} because in their lower-resolution 2DFS it partially overlaps the beat feature which has phase modulation.

Many pulsars with amplitude modulation are thought to be associated with nulling and/or mode-changing, while in other cases higher quality observations reveal weak drift.
In some pulsars nulling is found to be periodic, as observed in PSRs~B1133$+$16 \citep[$33P$;][]{hr07}, J1819$+$1305 \citep[$57P$;][]{rw08}, J0452$-$3418 \citep[$42P$;][]{gbm+24}, J1701$-$3726 \citep[$48P$;][]{wwd+23}.
Furthermore, PSRs~J1701$-$3726 and J2313$+$4253 are observed to have an additional fast periodic subpulse modulation similar to a flicker.
In some cases periodic nulling has been argued to arise from a sparsely filled rotating carousel, with ``pseudo nulls'' arising from a line of sight intersecting a region without active sub-beams \citep{hr07}.
However, as pointed out in \sect{sec:J1514-ps}, the nulling fraction of PSR~J1514$-$4834 is extremely low and unrelated to the flicker, and therefore must have different physical origins.

PSR~J1514$-$4834 is notable for the simultaneous presence of both amplitude modulation and drifting subpulses, a feature which has been reported for only a few pulsars: PSRs~B0943$+$10 \citep{dr99,dr01,ad01,gs03}, B0834$+$06 \citep{ad05}, B1857$-$26 \citep{mr08}, B1237$+$25 \citep{md14}, J1857$+$0057 \citep{yhw+23}, J0452$-$3418 \citep{gbm+24}, and J2022$+$5154 \citep{cww+24}.
All of these show a long period amplitude modulation in combination with fast drift.
PSR~J1514$-$4834 is the only pulsar known to exhibit fast amplitude modulation (flickers) alongside slow drift.
This will be argued to require a different interpretation (see \sect{sec:B0943}).
There are more pulsars for which \cite{sws+23} identified both amplitude modulation and drifting subpulses.
Although many of these may be related to mode changes rather than the simultaneous occurrence of distinct periodicities, some may potentially reveal common physical mechanisms or characteristics in similar systems.
Such findings would enhance our understanding of population-wide trends, taking us closer to a complete picture of the physics of the pulsar magnetospheres.

\subsection{The need for multiple underlying periodicities}
\label{sec:single-periodicity}

Here we will argue that a carousel, in its simplest form, is unable to explain the simultaneous appearance of the slow drift and flicker in PSR~J1514$-$4834. 

Regular drifting subpulses with $P_3$ close to, but not identical to $2$, will produce chessboard-like patterns in the pulse stack.
Such a pattern, at face value, looks like it contains a slow drift combined with alternating patterns.
This $P_3\approx2$ model has been applied to PSR~B0943$+$10 \cite[e.g.][]{bil18}, and a simulated example is shown in \fig{fig:Nyquist-simulation}.
As presented in \apdx{apdx:Nyquist-simulation}, the apparent slow drift arising in such a scenario is in fact a manifestation of a harmonic of the fundamental $P_3$ associated with the rapid drift. 
Therefore, if both the flicker and drifting subpulses observed in PSR~J1514$-$4834 are the consequence of a single carefully chosen carousel with $P_3 \approx 2$, it must require that the flicker feature in the 2DFS corresponds to the first harmonic (fundamental) and the drift feature corresponds to the aliased second harmonic\footnote{Note that there is no need to consider higher-order harmonics as their intensities decrease as harmonic order increases.}.
This implies that not only the intrinsic $1/P_3$ but also the $1/P_2$ frequency of the flicker feature should be half of that of the drift feature, which is clearly not the case (see \tab{tab:measure}).
More importantly, such a carousel with a single underlying periodicity fails to explain the presence of the beat features.

The presence of the beat features does require two frequencies which are at some level independent.
This is corroborated by the fact that the Fourier phases of the flicker and drifting subpulses are uncorrelated (see panel (c) of \fig{fig:J1514-phase-scatter}).
The two periodicities could be associated with completely independent processes. For example, periodic amplification of the radio emission in the magnetosphere modulating the drifting subpulse signal.
However, alternatively, the two periodicities might originate from a single integrated system, such as a more complex carousel model. Any two periodicities modulating each other will produce beat features, regardless of their physical origin. 
In \sect{sec:B0943}, it is explored if the circulation time of a carousel could be responsible for the additional frequency, which is followed by the discussion of an interpretation in the context of a feedback model (\sect{sec:feedback}). 

Two completely independent processes would be ruled out if correlated changes in the flicker and drifting subpulses can be established. Such a possible correlation is suggested in \sect{sec:J1514-fast}.
During the fast $P_3$ event, the pulse profile has a relatively strong leading component, which stands out significantly from random pulse-to-pulse variability.
Furthermore, this event coincides with an increase in flicker strength.
This could point to a magnetospheric state change that simultaneously affects the drifting subpulses and flicker. However, it remains to be seen if the only rough time alignment of these events is simply coincidence. Longer observations would establish this.

\subsection{Circulation time model}
\label{sec:B0943}

It has been shown that a carousel consisting of sub-beams with unequal intensities \citep{dr01} or non-uniform radial distances from the magnetic axis \citep{gs03} will result in simultaneous drifting subpulses plus amplitude modulation.
In such a scenario, the frequency of the amplitude modulation is set by the circulation time of the carousel.
Such a set-up is explored in \apdx{apdx:sidebands-simulation}, where it is illustrated that such a model indeed produces beat features.
Moreover, it is shown that the feature corresponding to the circulation time of a carousel with $n_{\rm sb}$ sub-beams has intrinsic (not aliased) $1/P_2$ and $1/P_3$ frequencies both of which are $n_{\rm sb}$ times smaller than those of the drift feature (see \fig{fig:sidebands-simulation}).
Therefore the feature associated with the circulation should have phase drift associated with it, although it could be mostly amplitude modulation if the number of sub-beams $n_{\rm sb}$ is large.

Here we will apply such a circulation time model to PSR~J1514$-$4834 and conclude that it cannot explain the observed fluctuation frequencies of the drift and flicker features.
In such a model, the circulation time is expected to be associated with the spectral feature which is mostly amplitude modulation, i.e. the flicker feature in the case of PSR~J1514$-$4834.
So unlike for PSR~B0943$+$10 and other pulsars in which this circulation time model has been applied, the feature associated with the circulation time has a higher observed $1/P_3$ frequency compared to that of the drifting subpulses.
Since the circulation time should intrinsically correspond to a lower $1/P_3$ frequency, at least the drift feature must be aliased.

The flicker feature observed in PSR~J1514$-$4834 has a $1/P_3$ frequency of $\approx0.49~\rm cpp$ (see \tab{tab:measure}) during the event where the measurements are most accurate.
Assuming its observed periodicity is unaliased, the circulation time would correspond to $P_4\approx2.04$.
This implies an exceptionally fast-rotating carousel, with a circulation time much shorter than predicted by the \citetalias{rs75} model ($P_4\approx17.8$).
If the observed flicker frequency were to be aliased, it would require an even faster rotating carousel.
Therefore, in what follows, we will only consider the case where the flicker frequency is not aliased.
On the other hand, the observed drifting subpulses have an observed $P_3$ significantly larger than the circulation time.
Yet, the expectation is that the $P_3$ should intrinsically be $n_{\rm sb}$ times smaller.
Consequently, the observed $1/P_3 \approx 1/24=0.04~\rm cpp$ must be an alias of a higher intrinsic frequency.

The alias of twice the flicker $1/P_3$ frequency is within the measurement uncertainties equal to frequency of the drifting subpulses.
So $n_{\rm sb}=2$ is compatible with the $1/P_3$ frequencies. However, this is incompatible with their $1/P_2$ frequencies.
This is because in the circulation time model both the $1/P_3$ and $1/P_2$ frequencies 
must be harmonically related (see \apdx{apdx:sidebands-simulation}).
Since there is no detection of phase modulation for the flicker feature, a large number of sub-beams $n_{\rm sb}$ is required. Hence the drift feature needs to be a higher-order harmonic of the flicker frequency.

The $1/P_2$ frequency of the drift feature should be  $n_{\rm sb}$ times larger than the flicker feature. 
For $n_{\rm sb}\geq4$ the $1/P_2$ frequency of the drift feature, predicted using the limit on how non-zero the $1/P_2$ frequency of the flicker frequency can be, is within the errors consistent with the measurement of the drift $1/P_2$ frequency. 
However, the drift feature being a fourth or higher-order harmonic of the flicker frequency is problematic. 

First of all, this would imply a high alias order for the drifting subpulse feature.
This would be at odds with \cite{sws+23} who concluded that the majority of pulsars exhibit, at most, a first-order alias.
More importantly, even slight deviations in the circulation time (fundamental) will result in changes in the apparent observed  $1/P_3$ frequency of the drift feature (higher-order harmonic) that are at least $n_{\rm sb}$ times larger. 
So if the intrinsic $1/P_3$ frequency of the drift feature is four times larger, also its spectral width should be four times larger than that of the flicker feature. 
Yet, especially in the long UHF-band observation, the width of the drift feature is narrower (panel (b) of \fig{fig:J1514-4834_2dfs}).
Therefore, for PSR~J1514$-$4834, the circulation time model is ruled out.

Since the circulation time model does not apply to PSR~J1514$-$4834, a different model is required. One possibility is explored in \sect{sec:feedback}. 
Such an alternative model might equally well apply to PSR~B0943$+$10. 
However, it remains to be seen how comparable these two pulsars are. 
Although the circulation time model, which relies on beats, was developed to explain PSR~B0943$+$10, the beat features (sidebands) and the corresponding amplitude modulation feature have never been simultaneously detected in one single observation \citep{dr99, dr01, bmr10, ad01, bil18}.
To avoid this dilemma, \cite{gs03} argue that a non-uniformly distributed sub-beams with random fluctuations could help explaining why the spectral feature associated with the circulation time is suppressed.

\subsection{Feedback model}
\label{sec:feedback}

As concluded in the previous subsection, the turning of a rigid near-unchanging carousel cannot explain the emission patterns of PSR~J1514$-$4834. 
Here we explore further if the flicker and drift periodicities could originate from a single integrated system. 
If the carousel picture is to be retained, it is required that as the individual sub-beams turn on the carousel they are subject to a separate but regular modulation process, resulting in the observed flickering patterns. 

To provide a possible interpretation of this extraordinary behaviour we may turn to a feedback model \citep{wri22} which suggests that all pulsars possess an inner carousel (often unseen, as in PSRs~J1514$-$4834 and B0943$+$10) that interacts over a time delay with the observed outer carousel.
The time delay then beats with the $\boldsymbol{E} \mathbf{\times} \boldsymbol{B}$ drift to create the double modulation effect in a single feedback system.
The idea of two carousels present in the magnetosphere is not far-fetched.
Over the intervening 50 years since its publication, the original carousel model \citep{rs75} has been modified, largely on observational grounds, to include a second carousel. 
This is required both by the presence of multiple components of integrated profiles \citep{ran83} and complex yet strongly linked subpulse patterns often found in all components of complex pulse profiles \cite[e.g.][]{bgg09}.
No more than two carousels have ever been proposed to explain observations, yet no physical argument has been proposed as to why more should not be formed on the polar cap. 

\begin{figure}
 \includegraphics[width=0.33\linewidth]{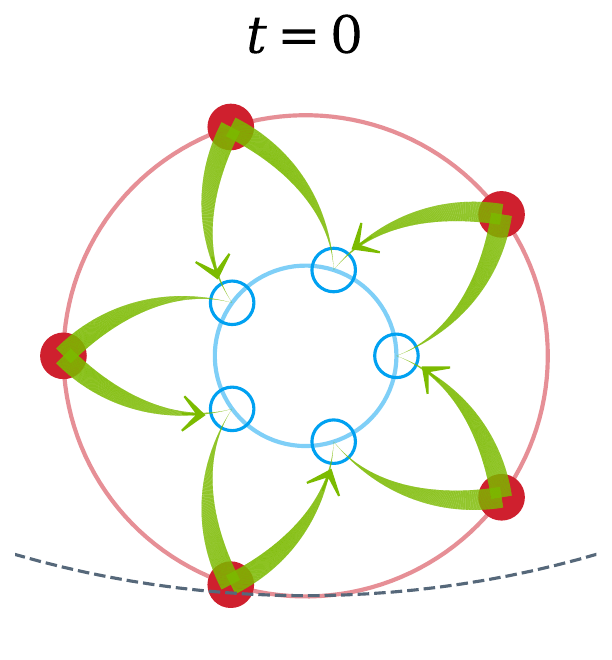}
 \includegraphics[width=0.33\linewidth]{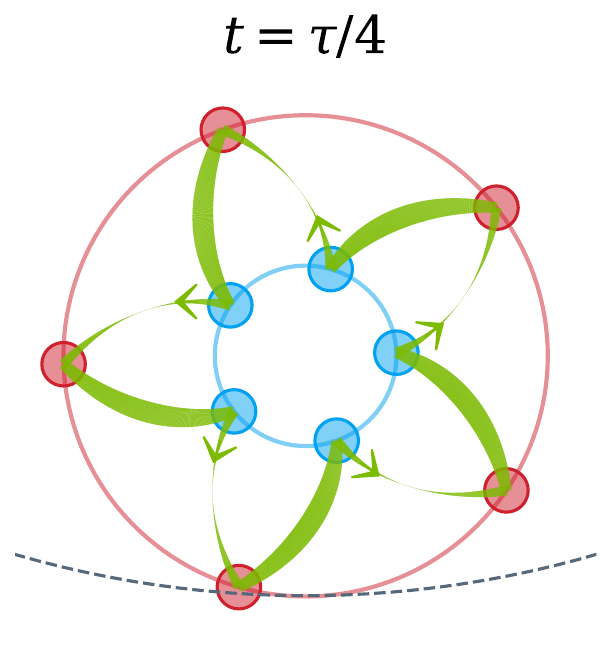}
 \includegraphics[width=0.33\linewidth]{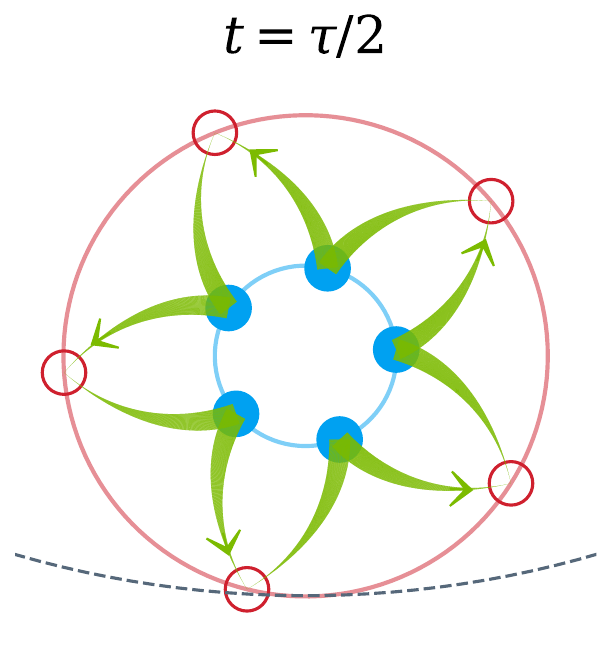}
  \caption{Geometric representations of a feedback system between inner (blue) and outer (red) carousels. 
  Here, the line of sight (curved dashed lines) only intersects the edge of the outer carousel.
  The width of the green line represents the strength of the interaction between the two carousels.
  The footprints, circles on the carousels, are active (filled) or inactive (open) depending on the strength of the interaction.
  Here shows three snapshots from the animation available on Zenodo at doi: \href{https://doi.org/10.5281/zenodo.14216938}{10.5281/zenodo.14216938}.
  }
  \label{fig:feedback}
\end{figure}

A geometric representation of the feedback model is presented in \fig{fig:feedback}, showing snapshots from a simulated animation at three time points.
In this illustrative simulation, five sparks are formed in both the inner (blue) and outer (red) carousels.
The interaction between two carousels in the form of an exchange of particle flows is represented by the curved green lines, which travel back and forth between the two carousels with a combined period of $\tau$.
The interaction of this flow (where the green line is thick) causes sparks in one carousel to seed localised sparks in the other carousel and vice versa, presumably via the magnetosphere.
Thus sparks are the "footprints" left by the particle flows travelling between two carousels.
In a steady state, the flow can be seen as a combination of a $\boldsymbol{E} \mathbf{\times} \boldsymbol{B}$ drift caused by an underlying magnetospheric circulation of fixed period $P_5$, and a quasi-radial movement between the inner and outer carousels. 

As a result, the {\it observed} circulation of the sparks is purely a mathematical consequence of the interplay (beat) between $P_5$ and $\tau$, and no longer the direct result of the $\boldsymbol{E} \mathbf{\times} \boldsymbol{B}$ drift as assumed in the \citetalias{rs75} model.
By adjusting the values of $P_5$ and $\tau$, it is possible to achieve a carousel that rotates in either direction.
For example, in \fig{fig:feedback}, both the $\boldsymbol{E} \mathbf{\times} \boldsymbol{B}$ drift and the spark circulation (both inner and outer carousels) are counterclockwise, which can be more easily confirmed in the animation (see Zenodo link in \fig{fig:feedback} caption).
For many pulsars this may provide advantages in explaining the observed drifting rates and directions, several of which are incompatible with the $\boldsymbol{E} \mathbf{\times} \boldsymbol{B}$ prediction \cite[e.g.][]{vsr+03, vt12, bmm+16, sws+23}.

As pointed out in \cite{wri22}, the flexibility of the feedback model arises from its capacity to specify the nature of the interaction between the two carousels. 
In our demonstration (\fig{fig:feedback}), this is visualised as the curved green lines of varying widths corresponding to varying strength.
Here a scenario is illustrated in which the intensity varies from strong to weak over a duration of $\tau/2$, followed by a further $\tau/2$ during which the interaction gradually increases back to strong and the cycle is complete.
As a consequence, the outer carousel (red) is brightest at $t=0$ (left panel), while the inner carousel (blue) is brightest at $t=\tau/2$ (right panel).
Which of the two carousels has brighter/weaker sparks would therefore alternate.
Since the line of sight grazes only the outer carousel (red), when the outer carousel is weak, little radio emission is observed, even though the inner carousel is bright.
This generates amplitude modulation in addition to the gradual slower drift of the sparks.

In general, the additional modulation pattern generated by the carousel interaction can vary over time in a single pulsar and/or from pulsar to pulsar.
For example, in the case of constant and smooth interaction, (i.e. uniform green lines occupying the entire flow), both carousels will always appear equally active, and drifting subpulses without additional modulation will be observed.
However, if gaps exist in the flow, this can lead to periodic nulling as observed, for example, in PSRs~B1133$+$16 \citep{hr07} and J1819$+$1305 \citep{rw08}.
If the strength of the interaction varies gradually, as in \fig{fig:feedback} rather than being abruptly turned on and off, the pulse energies will be modulated smoothly.
This can explain the continuous unimodal pulse-energy distribution of PSR~J1514$-$4834 (see \fig{fig:J1514-energy}).

For flickers to generate a pure amplitude modulation, as observed in PSR~J1514$-$4834, all sparks in the outer carousel must be synchronised such that they brighten and fade simultaneously.
This requires a specific pattern along the particle flows. For example, in \fig{fig:feedback} where five sparks are presented, five identical green curved lines are required to cover a full circulation. At $t=0$, each of them originates from the inner carousel (blue) as thin green line, gradually widens until reaching the outer carousel (red), then contracts and returns to the inner carousel before seamlessly connecting to the next green curved line.
So the maxima and minima of the flows reach all sparks of the same carousel at the same time.
However, it is possible to drop the assumption of a precise periodicity and allow a different number of green line segments, or a variation of the green lines in intensity, length or both.
Then the feedback system becomes capable of generating more complex drift patterns, and might explain the subtle effects found in pulsars such as PSRs~J1857$+$0057 \citep{yhw+23} and J1059$-$5742 \citep{sws+23}, where subpulse intensity appears to counter-drift from drift band to drift band.

In conclusion, the two periodicities observed for PSR~J1514$-$4834 can be explained in terms of a single integrated system which retains the concept of a carousel drift, but incorporates time-delayed interaction with an inner second carousel.
Such a feedback model provides a geometric (and possible physical) framework for understanding the link between the drifting subpulses and flickers observed in this pulsar.
It also holds out the possibility of understanding the complex emission of other pulsars which hitherto has been difficult to reconcile with the carousel model.

\section{Conclusion}
\label{sec:conclusion}

In this paper, we analyse the complex subpulse modulation patterns in PSR~J1514$-$4834 using L-band data from the Thousand-Pulsar-Array (TPA) programme and further MeerKAT UHF-band data.
In addition to drifting subpulses, PSR~J1514$-$4834 simultaneously shows rapid amplitude modulation with a period of $\approx2P$.
The latter is referred to as "flicker".
Such rapid amplitude modulation is uncommon in the pulsar population, and its coexistence with periodic drifting subpulses is even rarer (see \sect{sec:population}).
Also, PSR~J1514$-$4834 nulls extremely sporadically.
Its nulling fraction of just 0.03 per cent is as low as the lowest known in the pulsar population.
The flicker shows a continuous smooth unimodal pulse-energy distribution, distinct from that of the nulls. Clearly the nulls are of a different physical origin than the rest of the emission.

The flickers modulate the drifting subpulse pattern, and as a consequence spectral beat features should exist.
We theorised that these beats are characterised by predictable frequencies and Fourier phases -- each the sum and difference of those of the two periodicities. 
It is demonstrated that during a rare event, in which the drift, flicker, and one of the beat features are distinctly identifiable in the power 2DFS with high significance, the observed beat frequency matches what is expected (see \sect{sec:J1514-2DFS}).
The second predicted beat feature cannot be identified in this same data because, due to aliasing, the two beat features happen to overlap in the spectral domain.
However, by applying a newly developed Fourier phase correlation technique the presence of both the expected beat features is established (see \sect{sec:J1514-phase}), and this technique is shown to be robust in the presence of fluctuations in the strength and stability of the spectral features.
Therefore, the complex nature of the fluctuation spectra of PSR~J1514$-$4834 can be fully attributed to just two underlying periodicities.

The coexistence of multiple underlying periodicities in the emission of PSR~J1514$-$4834 creates challenges to the often invoked carousel model to explain drifting subpulses.
A carousel consisting of unequal sub-beams, as proposed to explain the coexistence of drifting subpulses and amplitude modulation in PSR~B0943$+$10 and other pulsars, is shown to be inapplicable to PSR~J1514$-$4834 (see \sect{sec:B0943}).
However, the feedback model of \cite{wri22} possesses sufficient flexibility to accommodate the complexities of the emission of PSR~J1514$-$4834, while retaining the concept of a carousel and providing a single integrated system that links the two observed periodicities (see \sect{sec:feedback}).
The feedback system is governed by an interaction time-scale which beats with the $\boldsymbol{E} \mathbf{\times} \boldsymbol{B}$ drift.
This results in both drifting subpulses and flickers, along with the expected beat features, as observed in PSR~J1514$-$4834.

\section*{Acknowledgements}
The MeerKAT telescope is operated by the South African Radio Astronomy Observatory, which is a facility of the National Research Foundation, an agency of the Department of Science and Innovation.
Pulsar research at Jodrell Bank Centre for Astrophysics and Jodrell Bank Observatory is supported by a consolidated grant from the UK Science and Technology Facilities Council (STFC). 
JAH acknowledges funding from the STFC Doctoral Training studentship.
GW thanks the University of Manchester for Visitor status. 
MeerTime data are housed and processed on the OzSTAR supercomputer at Swinburne University of Technology.

%%%%%%%%%%%%%%%%%%%%%%%%%%%%%%%%%%%%%%%%%%%%%%%%%%
\section*{Data Availability}
Data underlying this article will be shared upon reasonable request to the corresponding author.


\bibliographystyle{mnras}
\bibliography{references}

%%%%%%%%%%%%%%%%%%%%%%%%%%%%%%%%%%%%%%%%%%%%%%%%%%

%%%%%%%%%%%%%%%%% APPENDICES %%%%%%%%%%%%%%%%%%%%%

\appendix

\section{Fourier technique in drifting sub-pulse studies} 

\subsection{Two-dimensional Fourier spectrum}
\label{apdx:2DFT}

Following \cite{es02}, an idealised drifting subpulse pattern characterised by $P_2$ and $P_3$ in a pulse stack can be described as a product of a 2D sinusoid and an intensity envelope $I(\phi)$\footnote{This is Eq. (5) in \cite{es02} but without a phase envelope and without emission other than drifting subpulses.
The negative sign in front of $P_2$ is to ensure that a positive $P_2$ corresponds to a positive drift.}:
\begin{equation}
S(\phi,n)=I(\phi)\left[\cos\left(2\pi\left(-\frac{\phi}{P_2}+\frac{n}{P_3}\right)+\theta_0\right)+1\right].
\label{eq:2d-cos}
\end{equation}
Here $\phi$ is the rotational phase, corresponding to the pulse longitude from 0 to $360^\circ$, $n$ is the pulse number, and $\theta_0$ is the initial phase of the drifting subpulses relative to the start of the pulse stack.
It is the quantity $\theta_0$ we aim to quantify using the Fourier phase analysis.

\begin{figure*}
 \includegraphics[width=\textwidth]{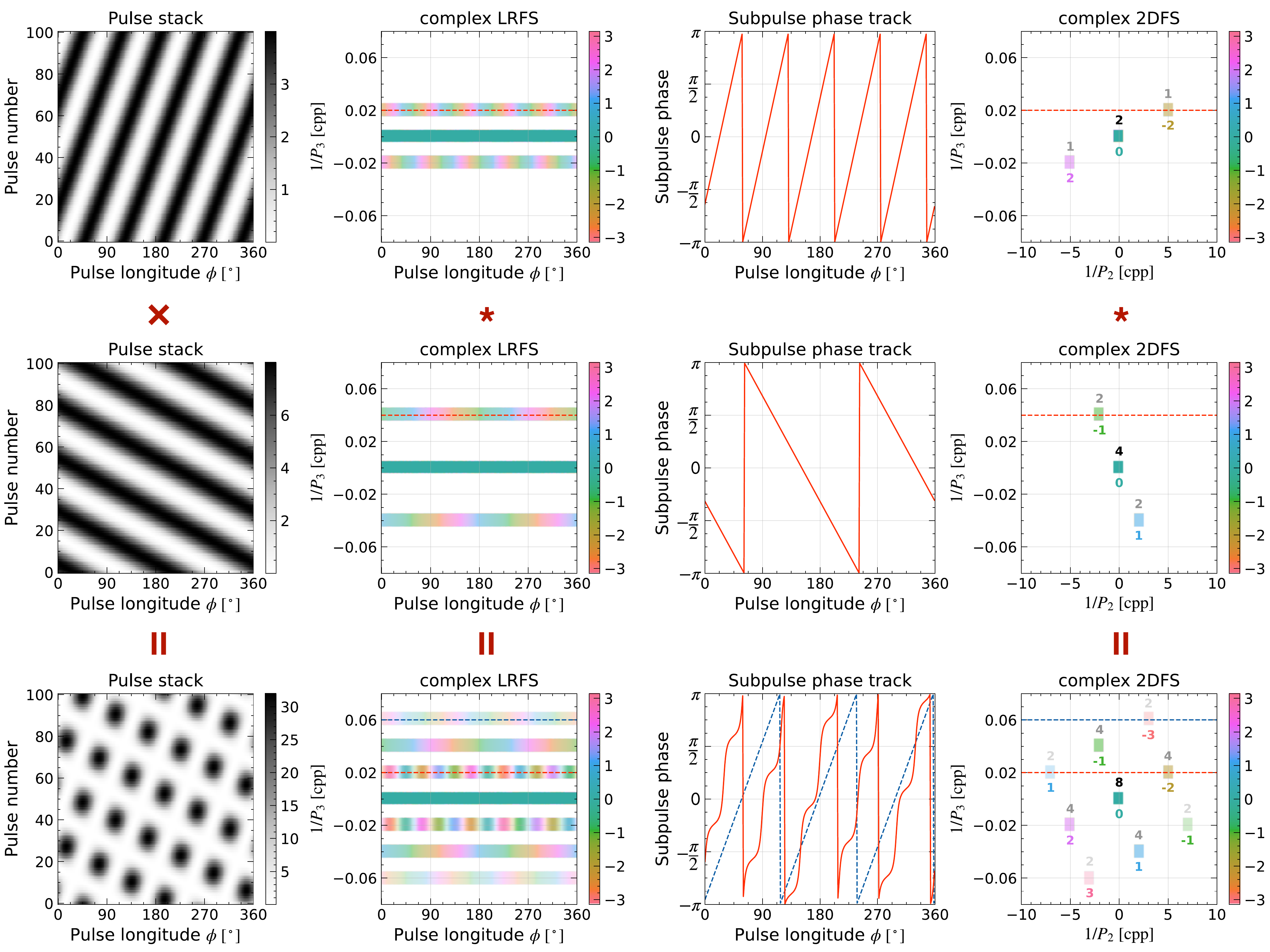}
  \caption{The top and middle rows show two 2D sinusoidal drifting subpulse signals with constant intensity envelopes, as well as their Fourier spectra. From left to right: the complex LRFS, subpulse phase tracks, and complex 2DFS.
  The product of these two drifting subpulse signals is presented in the bottom row, as well as its Fourier spectra.
  In the complex LRFS and 2DFS, colours indicate complex phases ranging from $\pi$ to $\pi$, while the transparency of each pixel reflects its magnitude.
  More explicitly, in complex 2DFS, the magnitude and phase values are annotated above and below the pixels containing the signal, respectively.
  The subpulse phase tracks correspond to the rows in the complex LRFS indicated by the dashed lines.
  }
  \label{fig:2DFT-2dsin-conv}
\end{figure*}

The 2D Fourier Transform of a pulse stack results in a 2D array comprising complex numbers. In the case of the idealised drifting subpulse signal given by \equ{eq:2d-cos}, the resulting spectrum\footnote{The definition of the 2D Fourier Transform used here differs from the conventional mathematical definition by having an additional negative sign in front of $k$. 
This cancels out the extra negative sign in front of $P_2$ in \equ{eq:2d-cos}, ensuring once again that a positive $P_2$ corresponds to a spectral feature appearing at a positive $k$.} is
\begin{equation}
    \begin{aligned}
    \widehat S_{\rm 2D}(k, l) &= {\cal F}_{\rm 2D}\{S(\phi, n)\} = \iint_{\phi,n} S(\phi, n) \mathrm{e}^{-\mathrm{i} 2 \pi\left(-k\phi+ln\right)}\mathrm{d}{\phi}\mathrm{d}{n} \\
    &= \widehat I(k) \ast \left[ \delta(k,l) + \frac{1}{2}  \delta\left(k-\frac{1}{P_2}, l-\frac{1}{P_3}\right)\mathrm{e}^{\mathrm{i}\theta_0}  \right. \\
    &\qquad\qquad\qquad\quad + \left. \frac{1}{2}\delta\left(k+\frac{1}{P_2}, l+\frac{1}{P_3}\right)\mathrm{e}^{-\mathrm{i}\theta_0}\right].
    \end{aligned}
    \label{eq:2d-delta}
\end{equation}
Here $\mathrm{i}$ is the imaginary unit while $k$ and $l$ are the coordinate axes in the Fourier domain, corresponding to the frequency components along the pulse longitude ($1/P_2$) and pulse number ($1/P_3$) directions respectively. 
The notation $\,\widehat{}\,$ denotes the Fourier Transform.
As per the convolution theorem, the spectrum is the convolution between $\widehat I(k)$ and a combination of three Dirac delta functions. The first delta function peaks at $k=l=0$ (DC term) and the second at $(k,l)=(1/P_2,1/P_3)$.
Since the pulse stack is real-valued, Hermitian symmetry is obeyed: $\widehat S_{\rm 2D}(k, l)=\widehat S_{\rm 2D}^\ast(-k, -l)$ and a third delta function appears at $(k,l)=(-1/P_2,-1/P_3)$.
To avoid redundancy, often only the top half of the spectrum is shown (as in for example \fig{fig:J1514-4834_2dfs} in the main paper).

To visualise the complex spectrum $\widehat S_{\rm 2D}(k, l)$, the magnitude $\left|{\widehat S_{\rm 2D}(k, l)}\right|$ and argument $\theta(k, l)$ spectra can be plotted separately. For \equ{eq:2d-delta} they are described by
\begin{equation}
    \begin{aligned}
    \left| \widehat S_{\rm 2D}(k, l)\right|
    &= 
    \begin{cases}
     \widehat I(k), & \text{if } l=0, \\
     \frac{1}{2} \widehat I(k-1/P_2), & \text{if } l=1/P_3, \\
     \frac{1}{2} \widehat I(k+1/P_2), & \text{if } l=-1/P_3, \\
     0, & \text{otherwise},
    \end{cases}
    \end{aligned}
    \label{eq:mag-spec}
\end{equation}
and
\begin{equation}
    \begin{aligned}
    \theta(k, l) = \angle \widehat S_{\rm 2D}(k, l)
    &= 
    \begin{cases}
     \angle \widehat I(k), & \text{if } l=0, \\
     \angle \widehat I(k-1/P_2) + \theta_0, & \text{if } l=1/P_3, \\
     \angle \widehat I(k+1/P_2) - \theta_0, & \text{if } l=-1/P_3, \\
     \text{undefined}, & \text{otherwise},
    \end{cases}
    \end{aligned}
    \label{eq:phase-spec}
\end{equation}
where 
the notation $\angle$ is used to denote the argument of the complex number, such that $\angle \mathrm{e}^{\mathrm{i}\theta}=\theta$.
The square of the magnitude spectrum $\left|{\widehat S_{\rm 2D}(k, l)}\right|^2$ represents spectral power, known as the conventional 2D fluctuation spectrum (2DFS) which provides insights into $P_2$ and $P_3$ of spectral features.
The argument spectrum $\theta(k, l)$ corresponds to what we refer to as the {\it "Fourier phase spectrum"}, which contains information about the initial phase $\theta_0$ as defined in \equ{eq:2d-cos}. 

In fact, the Fourier phases here are equivalent to the vertical offset of the subpulse phase track $\vartheta(\phi)$, which are the complex phases of the LRFS at a specific $1/P_3$ frequency bin.
For a pure 2D sinusoidal signal given by \equ{eq:2d-cos}, its LRFS (the 1D Fourier Transform of the pulse stack along the pulse number direction) is
\begin{equation}
    \begin{aligned}
    \widehat S_{\rm 1D}(\phi, l) &= {\cal F}_{\rm 1D}\{S(\phi, n)\} = \int_{n} S(\phi, n) \mathrm{e}^{\mathrm{i} 2 \pi ln}\mathrm{d}{n} \\
    &= I(\phi) \left[ \delta(l) + \frac{1}{2}  \delta\left(l-\frac{1}{P_3}\right)\mathrm{e}^{\mathrm{i}\left(2\pi\frac{\phi}{P_2}+\theta_0\right)}  \right. \\
    &\qquad\qquad\qquad + \left. \frac{1}{2}\delta\left(l+\frac{1}{P_3}\right)\mathrm{e}^{-\mathrm{i}\left(2\pi\frac{\phi}{P_2}+\theta_0\right)}\right].
    \end{aligned}
    \label{eq:2d-delta-lrfs}
\end{equation}
Similar to \equ{eq:2d-delta}, the definition of the 1D Fourier Transform here differs from the conventional mathematical definition by having an opposite sign in the exponent to ensure that positive drift corresponds to subpulse tracks with a positive gradient.
Based on \equ{eq:2d-delta-lrfs}, the subpulse phase track $\vartheta(\phi)$ is
\begin{equation}
    \vartheta(\phi) = 2\pi\frac{\phi}{P_2}+\theta_0
    \label{eq:subpulse_phase_track}.
\end{equation}
Therefore, its gradient is determined by $P_2$ and the intercept $\theta_0$ corresponds to the subpulse phase at pulse longitude $\phi=0$. This is identical to the Fourier phase in the complex 2DFS.

The top two rows of panels in \fig{fig:2DFT-2dsin-conv} give examples of pulse stacks with pure 2D sinusoidal signals where $I(\phi)$ is taken to be constant.
The corresponding complex LRFS, subpulse phase tracks, and complex 2DFS are shown as well.
In all these complex spectra, colours represent complex phases (ranging from $-\pi$ to $\pi$), while the transparency represents their magnitude (square root of power).
According to \equ{eq:2d-delta-lrfs}, the power in the LRFS (the second column) is concentrated at three frequencies: the DC component and two drift features at $\pm1/P_3$. The complex phases in the $+1/P_3$ bin, i.e. the subpulse phase tracks, are reproduced in the third column, and are described by \equ{eq:subpulse_phase_track}.
In the complex 2DFS (the fourth column), since $\widehat I(k)\propto\delta(k)$, Eqs.~(\ref{eq:mag-spec}) and (\ref{eq:phase-spec}) simplify to a pair of complex conjugate delta functions along with a DC component.
Here, it can be confirmed that the Fourier phases in complex 2DFS are indeed the intercepts of the subpulse phase tracks.

\begin{figure}
\centering
 \includegraphics[width=0.9\linewidth]{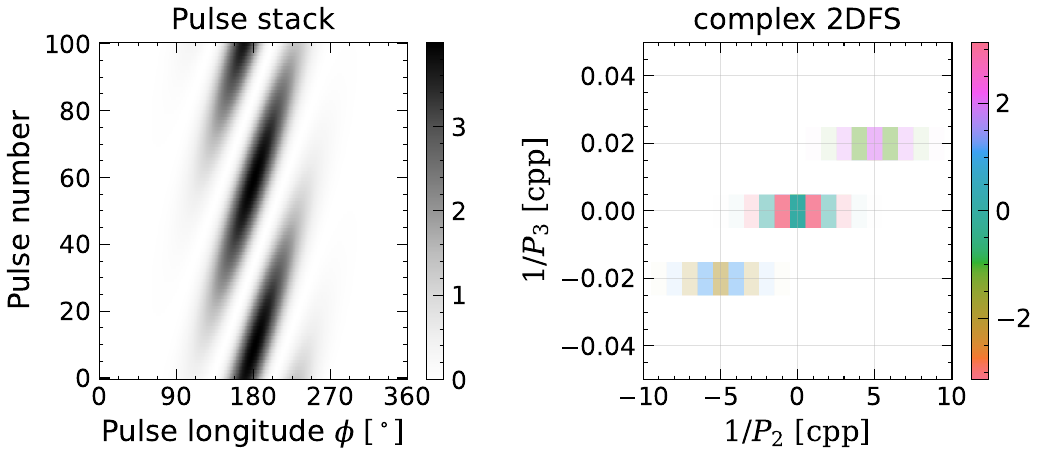}
  \caption{A pulse stack (left) of a 2D sinusoidal drifting subpulse signal with a Gaussian intensity envelope and its complex 2DFS (right). See \fig{fig:2DFT-2dsin-conv} for details.
  \label{fig:2DFT-2dsin-env}}
\end{figure}

However, in reality, the intensity envelope $I(\phi)$ will not be a constant function of the rotational phase $\phi$, and therefore a more realistic scenario involves a narrow on-pulse region.
For example, a non-constant and time-independent intensity envelope 
in the form of a Gaussian function is illustrated in \fig{fig:2DFT-2dsin-env}.
In such a scenario, the Fourier Transform of the windowed 2D sinusoid signal can be expressed as the convolution of the Fourier Transform of the window function (Gaussian envelope) and the delta functions representing the original (unwindowed) 2D sinusoid signal.
As a result, the spectral features are horizontally ``smeared out'' in the 2DFS and their phases exhibit rapid variability in the $1/P_2$ direction.
This complicates the measurement of the Fourier phase, which is crucial to identify an underlying beat system using the Fourier phase correlation technique introduced in \sect{sec:J1514-phase}.
We will show this explicitly in \sect{apdx:beat}.

\subsection{Beat system model}
\label{apdx:beat}

For a beat system, the drifting subpulse signal is modulated by another periodicity. As a result, beats arise.
Mathematically, the pulse stack $S$ is represented as a product of the drifting subpulse signal $S_{d}$ and the modulation pattern $S_{m}$:
\begin{equation}
    S = S_{d} \times S_{m}.
    \label{eq:driftxmod}
\end{equation}
In the following discussion, the subscripts $d$ and $m$ indicate quantities associated with the drift and modulation, respectively.
An illustrative example is presented in \fig{fig:2DFT-2dsin-conv}, where the pulse stacks, from top to bottom, correspond to $S_{d}$, $S_{m}$, and their product $S$.
According to the convolution theorem, the Fourier Transform of the pulse stack is
\begin{equation}
    \widehat S = \widehat S_{d} \ast \widehat S_{m}.
    \label{eq:drift*mod}
\end{equation}
If both $S_{d}$ and $S_{m}$ are pure sinusoidal modulations (e.g. described by \equ{eq:2d-cos}), we obtain (using \equ{eq:2d-delta})
\begin{equation}
    \widehat S(k,l) = \widehat I_{dm}(k) \ast \left( \delta(k,l) + \frac{H_d}{2} + \frac{H_m}{2} + \frac{H_{m+d}}{4} + \frac{H_{m-d}}{4} \right)
\label{eq:beat-delta}
\end{equation}
Here, we use the short notation 
\begin{equation}
    H_d = \delta(k-k_d,l-l_d)\mathrm{e}^{\mathrm{i}\theta_{d}} + \delta(k+k_d,l+l_d)\mathrm{e}^{-\mathrm{i}\theta_{d}},
\end{equation}
and 
\begin{equation}
\begin{aligned}
    H_{m \pm d} &= \delta(k-(k_m \pm k_d),l-(l_m \pm l_d))\mathrm{e}^{\mathrm{i}(\theta_{m} \pm \theta_{d})} \\
    &+ \delta(k+(k_m \pm k_d),l+(l_m \pm l_d))\mathrm{e}^{-\mathrm{i}(\theta_{m} \pm \theta_{d})},
\end{aligned}
\end{equation}
to represent pairs of Hermitian spectral features.
$H_m$ is defined similarly to $H_d$.
Moreover, we define $\widehat I_{dm}(k)=\widehat I_{d}(k) \ast \widehat I_{m}(k)$.
In \equ{eq:beat-delta}, the first term inside the brackets represents the DC component.
$H_d$ is attributed to the drifting subpulses and $H_m$ to the additional modulation.
The remaining two terms in the brackets, $H_{m+d}$ and $H_{m-d}$, are identified as beat features, which are the sum (labelled as $\Sigma$ in the following discussion and main text) and difference (labelled as $\Delta$) of the two input features in terms of not only the location of the delta function (frequency) but also the Fourier phase.
Because of the complex conjugate nature of Fourier spectra (Hermitian symmetry), each term except the DC component has both positive and negative frequency components.
Therefore, there are a total of nine spectral features, as shown in the bottom-right 2DFS in \fig{fig:2DFT-2dsin-conv}.

\equ{eq:beat-delta} can be expressed as follows in terms of magnitude and phase\footnote{These spectral features are assumed to be separated in the $l$ ($1/P_3$) direction and not overlap with each other in the spectra.}:
\begin{equation}
    |\widehat S(k, l)| = 
    \begin{cases} 
    | \widehat{I}_{dm}(k) | & \text{if } l = 0, \\
    | \widehat{I}_{dm}(k \mp k_d) | & \text{if } l = \pm l_d, \\
    | \widehat{I}_{dm}(k \mp k_m) | & \text{if } l = \pm l_m, \\
    | \widehat{I}_{dm}(k \mp k_\Sigma) | & \text{if } l = \pm l_\Sigma, \\
    | \widehat{I}_{dm}(k \mp k_\Delta) | & \text{if } l = \pm l_\Delta.
    \end{cases}
    \label{eq:mag-spec-beat}
\end{equation}
\begin{equation}
    \theta(k,l) = 
    \begin{cases} 
    \angle \widehat{I}_{dm}(k) & \text{if } l = 0, \\
    \angle \widehat{I}_{dm}(k \mp k_d) \pm \theta_d & \text{if } l = \pm l_d, \\
    \angle \widehat{I}_{dm}(k \mp k_m) \pm \theta_m & \text{if } l = \pm l_m, \\
    \angle \widehat{I}_{dm}(k \mp k_\Sigma) \pm \theta_\Sigma & \text{if } l = \pm l_\Sigma, \\
    \angle \widehat{I}_{dm}(k \mp k_\Delta) \pm \theta_\Delta & \text{if } l = \pm l_\Delta.
    \end{cases}
    \label{eq:phase-spec-beat}
\end{equation}
Here
\begin{equation}
\begin{array}{ccc}
    \begin{array}{l}
        \begin{aligned} 
          k_\Delta &= k_m - k_d, \\
          k_\Sigma &= k_m + k_d,
        \end{aligned}
    \end{array}
&
    \begin{array}{l}
        \begin{aligned} 
          l_\Delta &= l_m - l_d, \\
          l_\Sigma &= l_m + l_d,
        \end{aligned}
    \end{array}
&
    \begin{array}{l}
        \begin{aligned} 
            \theta_\Delta &= \theta_m - \theta_d, \\
            \theta_\Sigma &= \theta_m + \theta_d.
        \end{aligned}
    \end{array}
\end{array}
\label{eq:mag-phase-prediction}
\end{equation}
In the case of constant intensity envelopes (\fig{fig:2DFT-2dsin-conv}), i.e. $I_{d}(\phi)$ and $I_{m}(\phi)$ are constant, $\widehat I_{dm}(k)$ will be a delta function and therefore 
all $\angle \widehat I_{dm}(k)$ terms in \equ{eq:phase-spec-beat} will be zero, leading to a concise conclusion: the $1/P_2$ and $1/P_3$ frequencies, as well as the Fourier phases of the beat features, are the sum and the difference of those of the drifting subpulses and the additional modulation.
This can be verified in the complex 2DFS in \fig{fig:2DFT-2dsin-conv}.
For example, the Fourier phases of the drifting subpulses in the top and middle row are $-2$ and $-1$ rad, respectively. The $\Sigma$ feature resulting from the beat of these two drifting subpulse patterns (the top-right pixel in the bottom-right 2DFS) has a Fourier phase of $-2+(-1)=-3$ rad.
Even in the presence of the dramatic phase variability in the $1/P_2$ direction arising if there is a non-constant intensity envelope (such as shown in \fig{fig:2DFT-2dsin-env}), the Fourier phases of the beat features can be shown to be correlated with the sum or difference of the Fourier phases of the two features presumed to be beating.
This will be confirmed in \apdx{apdx:WSRT}.
Consequently, measurements of the beat frequencies (their positions in the 2DFS, see \sect{sec:J1514-2DFS}), as well as their Fourier phases (see \sect{sec:J1514-phase}) can serve as an independent verification to identify a beat system as the cause of complex-looking spectra.

In principle, a beat system can also be identified based on subpulse phase tracks, given their mathematical similarity.
For instance, in \fig{fig:2DFT-2dsin-conv}, the Fourier phase of the $\Sigma$ feature (the top-right pixel in the bottom-right 2DFS) is $-3$ rad, corresponding to the intercept of the subpulse phase track (blue dashed line).
However, an advantage of using the complex 2DFS using the approach developed in this paper is to separately measure the Fourier phases of two spectral features with overlapping $1/P_3$ frequencies but different $1/P_2$ frequencies.
For example, in the bottom row of \fig{fig:2DFT-2dsin-conv}, two spectral features are at $1/P_3=0.02~\rm cpp$ in the 2DFS (indicated by the red horizontal line), and their Fourier phase information is mixed in the subpulse phase track (red curvy line), making the identification of an intercept ambiguous.
In the case of PSR~J1514$-$4834, for the UHF-band data in panel (c) of \fig{fig:J1514-4834_2dfs} in particular, both beat features ($\Sigma$ and $\Delta$) overlap with the flicker feature in terms of $1/P_3$, while $1/P_2$ is well separated.
For such a scenario, the Fourier phases in the complex 2DFS provide a more effective approach to analyse the phase information of the beat and flicker features.
In fact, even if spectral features were not well separated, the Fourier phase correlation technique would still be viable (see \apdx{apdx:corr_other} and \fig{fig:J1514-4834_UHF_phase_alias}).

\section{Applications of Fourier phase correlation methodology}

\subsection{A synthetic beat system}
\label{apdx:WSRT}

To demonstrate the Fourier phase correlation methodology and its application, we construct a beat system using real pulsar data.
Since a pulse stack from a beat system can be thought of as a product of two pulse stacks characterised with different $P_2$ and $P_3$ periodicities (\equ{eq:driftxmod}), we multiply two pulse stacks of real pulsar data to create synthetic beat system data.
Therefore, all the complexities of realistic pulsar data not captured by the simplified mathematical model as used in \fig{fig:2DFT-2dsin-conv} are included.

\begin{figure*}
 \includegraphics[width=0.8\textwidth]{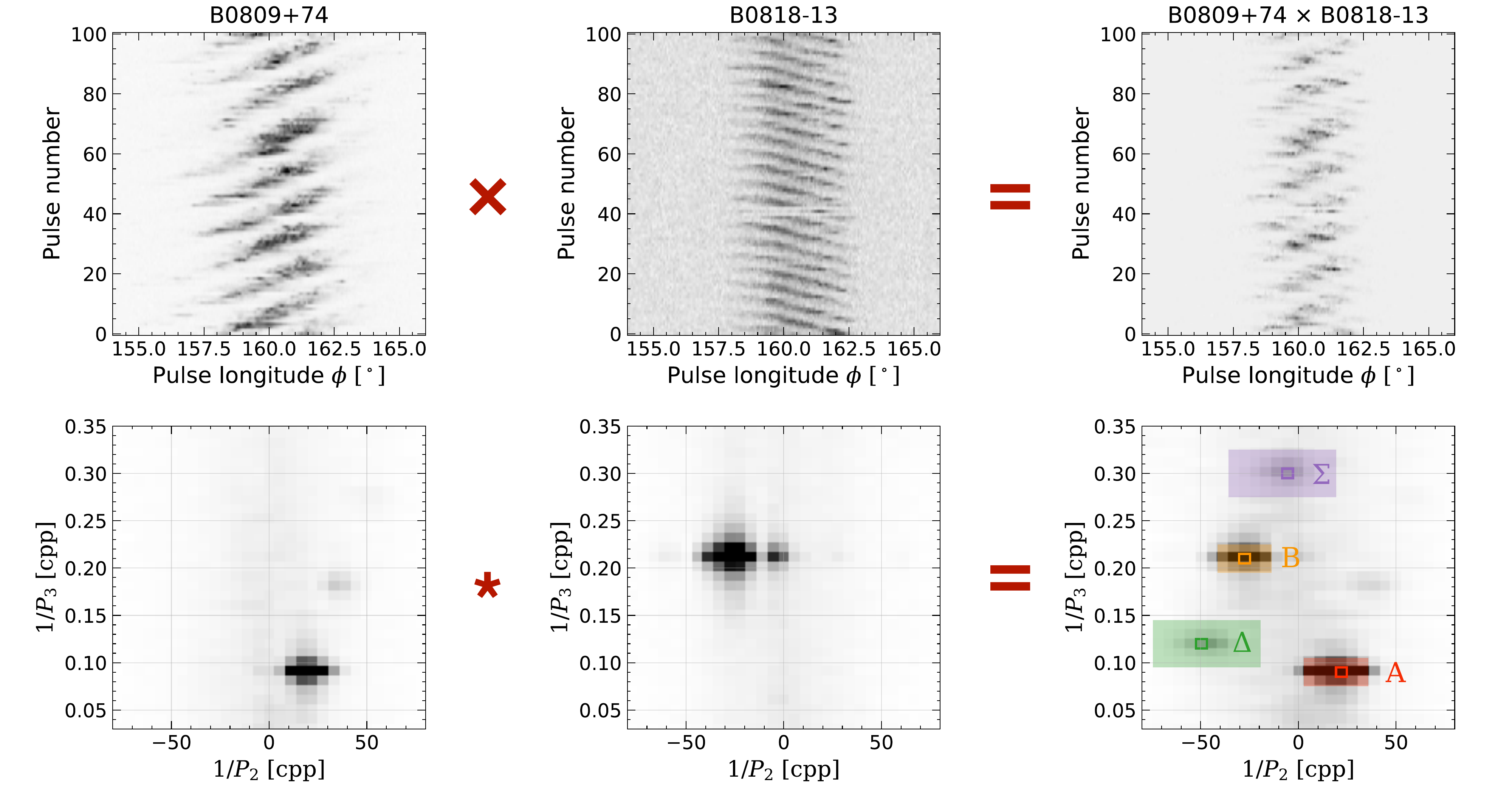}
  \caption{The synthetic beat system constructed from 8,600 pulses of PSRs~B0809$+$74 and B0818$-$13. Only the first 100 pulses are displayed in the pulse stacks in the top row.
  To highlight the weaker features, the brighter features are clipped by setting the threshold to 50 per cent of the maximum power in each panel.
  The spectral features are labelled in a way similar to panel (c) of \fig{fig:J1514-4834_2dfs}.}
  \label{fig:B0809_B0818_flip}
\end{figure*}

The synthetic data is shown in \fig{fig:B0809_B0818_flip}, where the pulse stacks of two pulsars, PSRs~B0809$+$74\footnote{The pulse stack of PSR~B0809$+$74 is horizontally flipped to obtain a positive $P_2$. This is done to maximise the separation of all features in the resulting spectrum and minimize overlap among them.} and B0818$-$13, are multiplied.
This data is the same as published in \cite{wel07} and were obtained using the Westerbork Synthesis Radio Telescope (WSRT) at a centre frequency of 1380 MHz.
In this observation, PSR~B0809$+$74 has a $P_3=11.12$ and a $P_2=-13.2^\circ$, while PSR~B0818$-$13 has a $P_3=4.74$ and a $P_2=-5.1^\circ$.
Sections of their pulse stacks are displayed in the upper row in \fig{fig:B0809_B0818_flip}, and the corresponding power 2DFS are shown in the bottom row. The right-hand side pulse stack is the product of the pulse stacks from two pulsars, hence shows beats.
As a result, in addition to the two spectral features attributed to the drifting subpulses of PSRs~B0809$+$74 (labelled as $A$) and B0818$-$13 (labelled as $B$), two further features ($\Sigma$ and $\Delta$) are generated in the far-right 2DFS. Their positions in the spectrum are indeed the sum and difference of the frequencies these two pulsars have as predicted by \equ{eq:mag-phase-prediction}.

\begin{figure}
    \includegraphics[width=\linewidth]{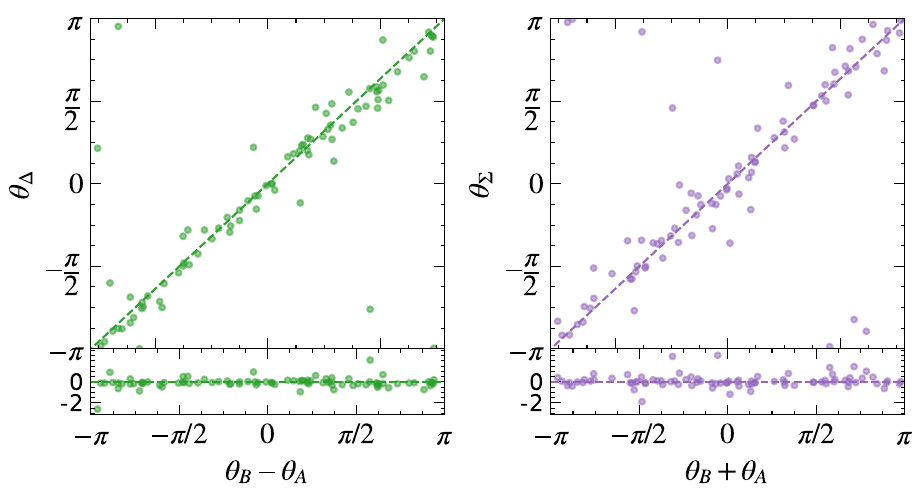}
     \caption{The Fourier phases of the pixels indicated in \fig{fig:B0809_B0818_flip} are plotted in a way similar to \fig{fig:J1514-phase-scatter}. The circular correlation coefficients $r$ are 0.83 (left) and 0.74 (right).}
     \label{fig:B0809_B0818_phase}
   \end{figure}

Following the procedures outlined in \sect{sec:J1514-phase}, we can extract the Fourier phases of the spectral features in the complex 2DFS and compare them to the predictions from the beat system model. 
The results are presented in \fig{fig:B0809_B0818_phase} in a way similar to \fig{fig:J1514-phase-scatter}.
The highly significant correlations indicate that the Fourier phases of the beat features are indeed the sum and difference of those of the two features beating, as expected.

\begin{figure}
\centering
    \includegraphics[width=0.9\linewidth]{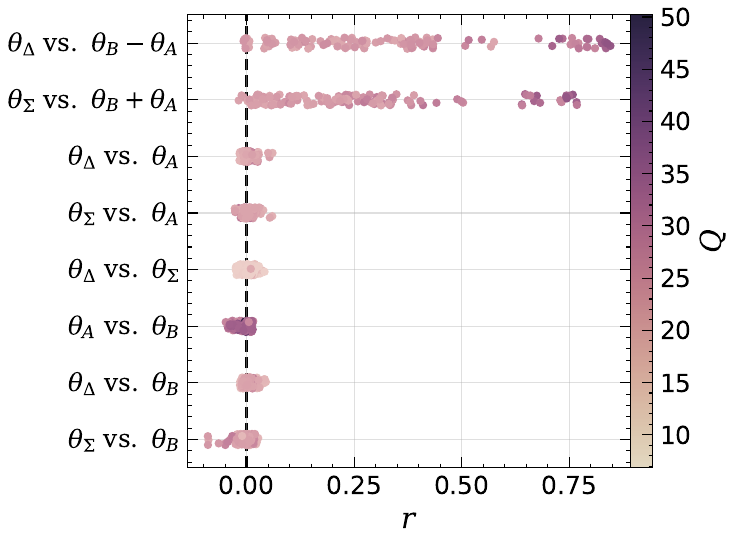}
     \caption{The distribution of the circular correlation coefficients $r$ by selecting different pixels in the shaded regions in \fig{fig:B0809_B0818_flip}.
     The colours of the dots represent the signal strength $Q$ of the pixels used to calculate the Fourier phase correlation (see \apdx{apdx:WSRT}). For clarity, only the top half of pixel combinations based on $Q$ in each group are shown.}
     \label{fig:B0809_B0818_phase_all_r_dist}
   \end{figure}

Considering that the spectral features are broadened, we repeat the above procedure by selecting different pixels in the shaded regions in \fig{fig:B0809_B0818_flip}.
For each pixel combination, a corresponding circular correlation coefficient $r$ is obtained (as defined in \apdx{apdx:r_fl}), which is then used to build the distributions presented in \fig{fig:B0809_B0818_phase_all_r_dist}, where all the spectral features are compared with each other.
It is clear that in most of the pixel combinations, the positive correlations between $\theta_\Delta$ and $\theta_B-\theta_A$ (as well as $\theta_\Sigma$ and $\theta_B+\theta_A$) are observed.
The significance of this is highlighted by comparing it with the absence of correlations found for other spectral features which are not expected to be correlated.
For example, the Fourier phases for features $A$ and $B$ are completely uncorrelated (labelled $\theta_B$ vs. $\theta_A$), as must be true given they were generated by two different pulsars.

In some pixel combinations, we observe a relatively low correlation between $\theta_\Delta$ ($\theta_\Sigma$) and $\theta_B-\theta_A$ ($\theta_B+\theta_A$).
This is because in those pixels, the signal is weak.
To quantify this effect, the signal strength $Q$ is calculated as follows for cases involving only two pixels:
\begin{equation}
    Q = \frac{\sqrt{(M_1-\mu_\mathrm{off})(M_2-\mu_\mathrm{off})}}{\sigma_\mathrm{off}}.
\end{equation}
Here $M_1$ and $M_2$ are the complex magnitudes of the two pixels, and $\mu_\mathrm{off}$ and $\sigma_\mathrm{off}$ are the mean and standard deviation of the magnitude of all pixels not associated with spectral features.
If three pixels are involved, the process is similar, except that the cube root of the product of three terms are used. This $Q$ is a proxy for the effective signal-to-noise ratio of the pixels selected and is used to colour the points in \fig{fig:B0809_B0818_phase_all_r_dist}.
Reviewing the distribution of $r$ between $\theta_\Delta$ ($\theta_\Sigma$) and $\theta_B-\theta_A$ ($\theta_B+\theta_A$) shows that points with relatively low correlation have light colours, corresponding to weaker signals.

\subsection{PSR~J1514$\texorpdfstring{-}{-}$4834}
\label{apdx:corr_other}

\begin{figure}
 \includegraphics[width=1\linewidth]{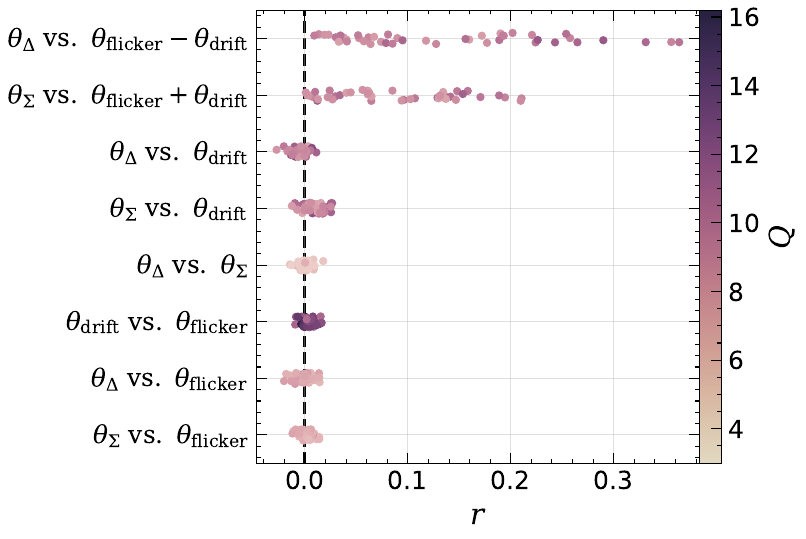} \\
 \includegraphics[width=1\linewidth]{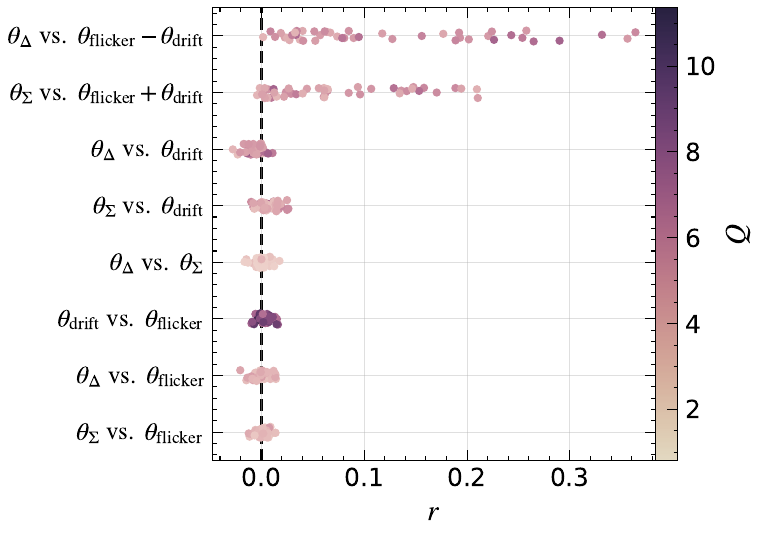}
  \caption{Similar to \fig{fig:B0809_B0818_phase_all_r_dist} but for PSR~J1514$-$4834 based on the pixels in the shaded regions in panel (c) of \fig{fig:J1514-4834_2dfs} using the UHF- (upper) and L-band (lower) data.}
  \label{fig:J1514-4834_UHF_phase_all_r_dist}
\end{figure}

The distribution of the circular correlation coefficients $r$ between the Fourier phases of each spectral feature of PSR~J1514$-$4834 observed in both the UHF- and L-band data is shown in \fig{fig:J1514-4834_UHF_phase_all_r_dist}, using the pixel combinations from the shaded regions in panel (c) of \fig{fig:J1514-4834_2dfs}.
Positive correlations are observed between $\theta_\Delta$ ($\theta_\Sigma$) and $\theta_{\rm flicker}-\theta_{\rm drift}$ ($\theta_{\rm flicker}+\theta_{\rm drift}$), while other combinations are uncorrelated. This follows the prediction for a beat system (\apdx{apdx:beat}).

\begin{figure*}
 \includegraphics[width=0.8\textwidth]{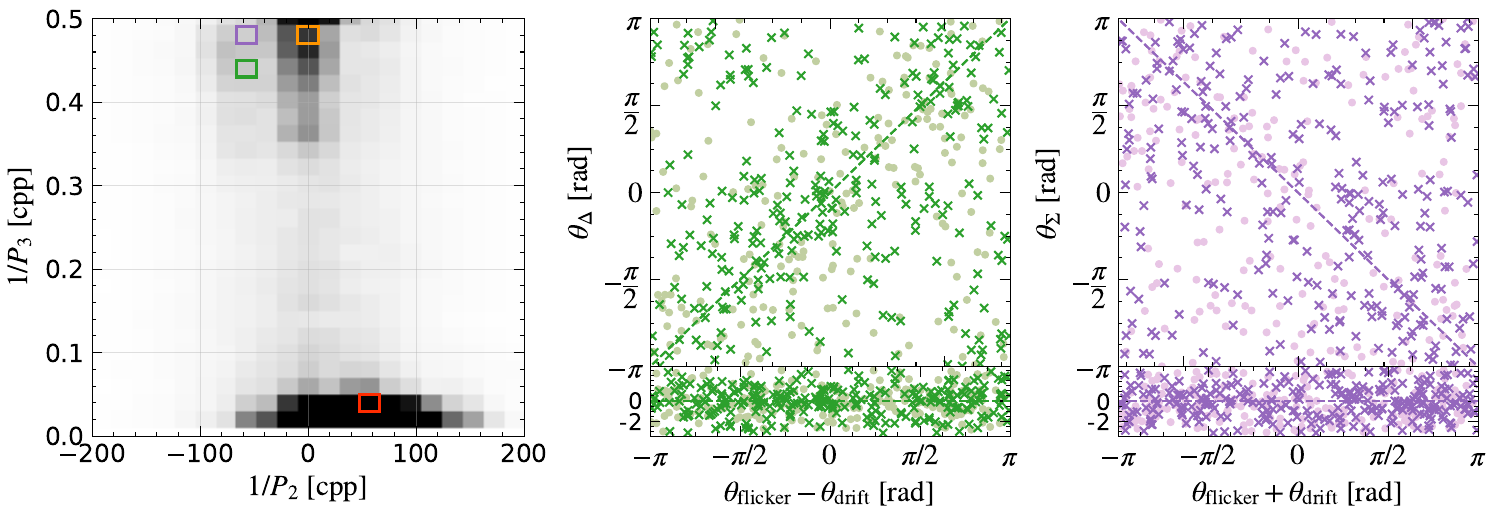} \\
 \includegraphics[width=0.8\textwidth]{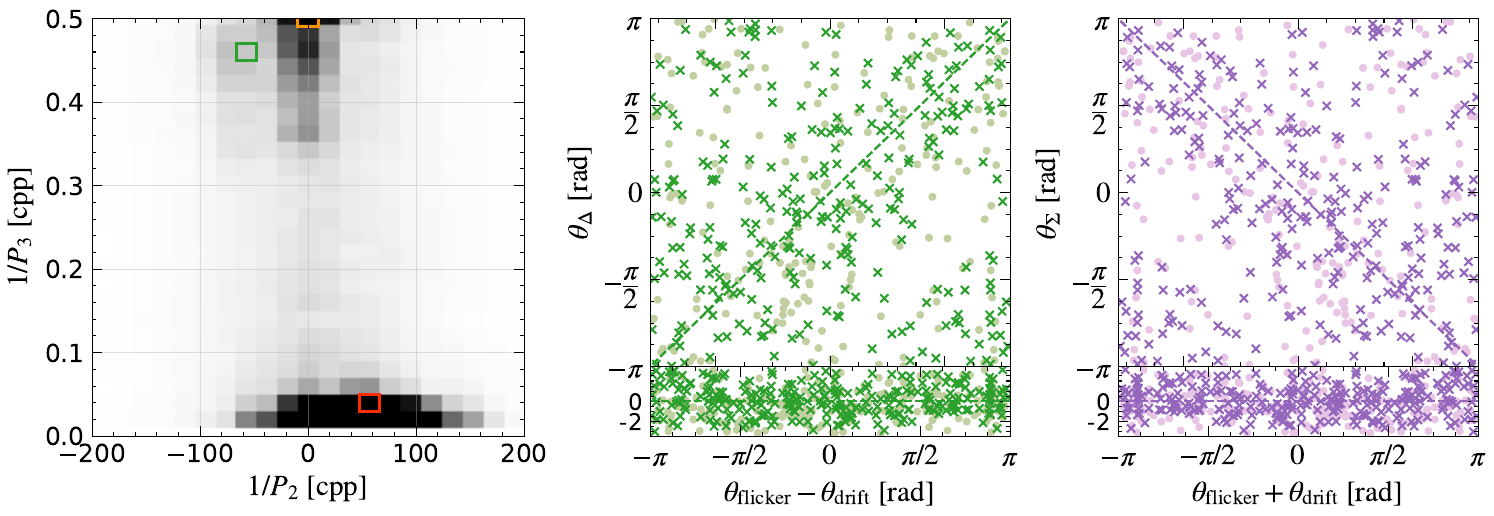} 
  \caption{The results from the Fourier phase correlation analysis for two pixel combinations for PSR~J1514$-$4834 where the $\Sigma$ beat feature is aliased. 
  The 2DFS in the left column are reproduced from \fig{fig:J1514-4834_2dfs}.
  The scatter plots are similar to those in \fig{fig:J1514-phase-scatter}.
  In the case shown in the upper row, the $\Sigma$ beat feature (purple) is aliased and appears on the other side of the 2DFS.
  Based on a combination of the UHF- (dots) and L- (crosses) band data, a positive correlation of $r=0.12~(p=5\times10^{-26})$ and a negative correlation of $r=-0.04~(p=3\times10^{-10})$ are obtained in the middle and right scatter plots, respectively.
  In the lower row, the flicker feature (orange) is chosen to be exactly at the Nyquist frequency ($0.5~\rm cpp$), so that the alias $\Sigma$ beat feature exactly overlaps with the $\Delta$ beat feature.
  The correlation coefficients have opposite signs with equal magnitude: $r=\pm0.06~(p=1\times10^{-13})$.}
  \label{fig:J1514-4834_UHF_phase_alias}
\end{figure*}

\fig{fig:J1514-4834_UHF_phase_alias} shows the Fourier phase correlation results for two individual pixel combinations in a way similar to Figs.~\ref{fig:J1514-4834_2dfs} and \ref{fig:J1514-phase-scatter}.
For both cases, $\theta_\Delta$ follow $\theta_{\rm ficker}-\theta_{\rm drift}$, leading to positive correlations shown in the middle column.
However, in both cases, the $\Sigma$ feature has a $1/P_3$ frequency exceeding the Nyquist frequency ($0.5~\rm cpp$), resulting in an aliased $\Sigma$ feature appearing on the other side of the 2DFS, as indicated by the purple outlined boxes in the 2DFS in \fig{fig:J1514-4834_UHF_phase_alias}.
The aliasing effect reverses the sign of $\theta_\Sigma$, leading to the expectation 
that $\theta_\Sigma$ should be anti-correlated with $\theta_{\rm ficker}+\theta_{\rm drift}$. 
Indeed a weak anti-correlation can be seen in the third column.
Note, in the second row of \fig{fig:J1514-4834_UHF_phase_alias}, the pixel in the flicker feature is chosen to be at $1/P_3=0.5~\rm cpp$ so that the corresponding $\Sigma$ and $\Delta$ beat features exactly overlap.
In this case, the circular correlation coefficients $r$ have the same magnitude but opposite signs (see caption of \fig{fig:J1514-4834_UHF_phase_alias}).
This shows that the Fourier phase correlation technique can be effective even under the influence of aliasing effects.
For clarity, pixel combinations involving aliased $\Sigma$ beat frequencies were excluded in \fig{fig:J1514-4834_UHF_phase_all_r_dist}.

\section{Circular correlation coefficient}
\label{apdx:r_fl}

The Pearson correlation coefficient $r_p$ is often used to quantify the correlation between two variables. However, circular data such as angles and phases are cyclic in nature.
Hence, to find correlations in the Fourier phases, the Pearson correlation coefficient $r_p$ is not appropriate and can be expected to underestimate the true correlation.

To account for cyclical continuity,
\cite{fl83} proposed a circular correlation coefficient, denoted as $r_{fl}$ here. The correlation between two circular variables $\theta$ and $\phi$ is defined to be
\begin{equation}
    r_{fl}=\frac{\sum\limits_{1 \leq i<j \leq n} \sin \left(\theta_i-\theta_j\right) \sin \left(\phi_i-\phi_j\right)}{\sqrt{\sum\limits_{1 \leq i<j \leq n} \sin ^2\left(\theta_i-\theta_j\right) \sum\limits_{1 \leq i<j \leq n} \sin ^2\left(\phi_i-\phi_j\right)}}.
\end{equation}
In an extreme case where the distributions of $\theta$ and $\phi$ are each unimodal and highly concentrated, $r_{fl} \approx r_p$.
The circular correlation coefficient $r_{fl}$ shares similar properties as the Pearson correlation coefficient $r_p$. For example,
\begin{enumerate}
    \item $-1 \le r_{fl} \le 1$;
    \item $r_{fl}=0$ if $\theta$ and $\phi$ are independent;
    \item $r_{fl}$ is invariant under choice of the origin for both variables;
    \item if $r_{fl}=r$ is measured between $\theta$ and $\phi$, then $r_{fl}=-r$ when measured between either $-\theta$ and $\phi$ or $\theta$ and $-\phi$.
\end{enumerate}

However, a significant difference between the circular correlation coefficient $r_{fl}$ and the Pearson correlation $r_p$ is that $r_{fl}=\pm1$ if and only if $\phi=\pm\theta+\rm constant$.
For example, in cases of $\phi=2\theta$, despite the linear relationship between $\theta$ and $\phi$, the circular correlation coefficient $r_{fl}$ would be zero.
This feature is particularly useful in our Fourier phase correlation analysis introduced in \sect{sec:J1514-phase} because the Fourier phase of a beat feature inherently equals the difference or sum of phases of the two spectral features beating.

Another advantage of adopting the circular correlation coefficient $r_{fl}$ in our analysis is the fact that arbitrary constant shifts of the phases do not affect the correlation, meaning that the circular correlation between two variables $\theta_i$ and $\phi_i$ is the same as that between $\theta_i$ and $\phi_i+\mathrm{constant}$.
Due to various factors, we may consistently overestimate or underestimate the Fourier phases of spectral features, introducing an intercept in the correlation scatter plots.
This will not affect the circular correlation coefficient $r_{fl}$, thereby adding robustness to the results.

To quantify the significance of a correlation, it should be tested if a measured $r_{fl}$ is incompatible with the null hypothesis that the measured $\theta_i$ and $\phi_i$ values are drawn from an underlying distribution with $r_{fl}=0$.
For pulsar data, $\theta$ and $\phi$ will have uniform distributions.
If uncorrelated, it can be shown that if $n$ samples of each are drawn randomly,  $nr_{fl}$ follows a Laplace distribution such that the probability density function of $nr_{fl}$ is $\frac12 e^{-|nr_{fl}|}$ \citep{fis93}.
Thus, for a one-sided test at the significance level $\alpha$, if measured $nr_{fl}>-\ln 2\alpha$, the null hypothesis that $r_{fl}=0$ is rejected and $\theta$ and $\phi$ are significantly correlated.

\section{Rotating carousel model simulations}

\subsection{Carousel with single periodicity}
\label{apdx:Nyquist-simulation}

\begin{figure}
    \centering
    \includegraphics[width=\linewidth]{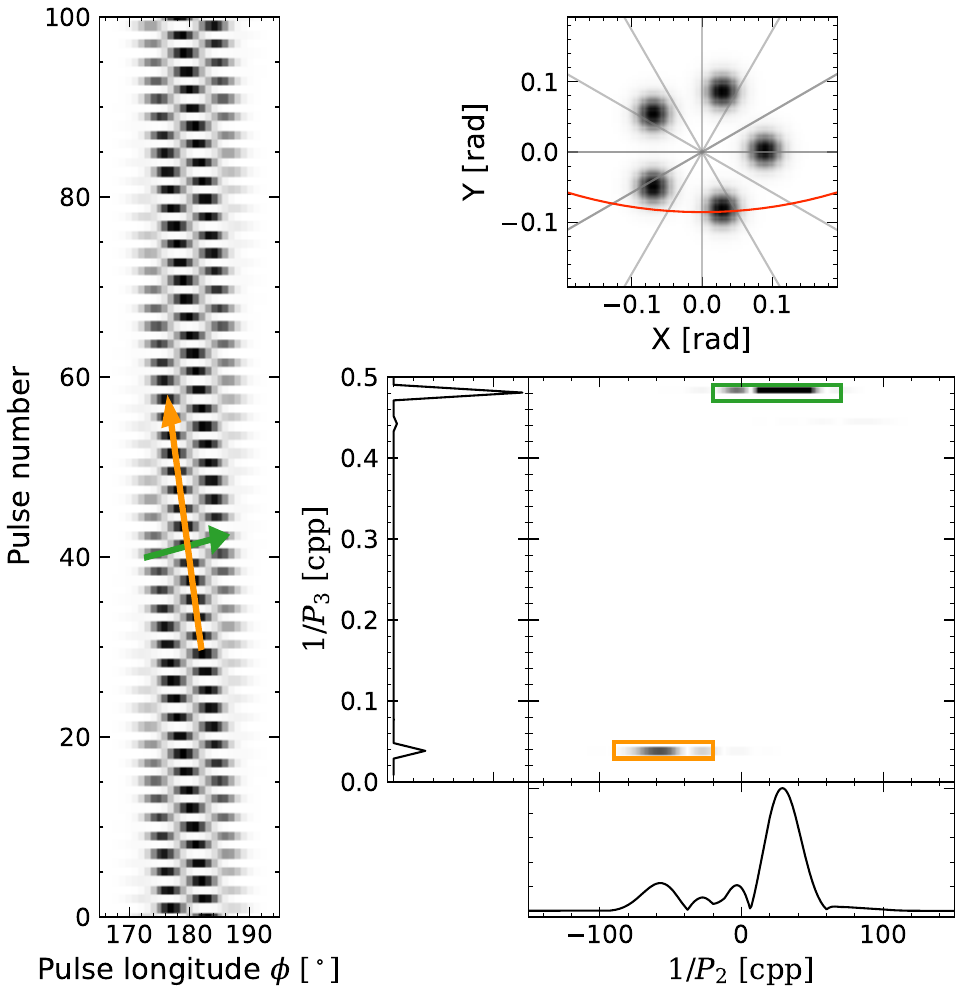}
    \caption{Simulated pulse stack (left) with a carousel consisting of five sub-beams and its 2DFS (bottom-right). The used carousel configuration is shown in the top-right panel, where the black spots represent the sub-beams. The red curved dashed line is part of the trace of the line of sight in the emission region. The magnetic axis is situated at the centre and the carousel rotates around this. Given the input circulation time $P_4$ and the number of sub-beams $n_{\rm sb}$, the resulting drift with $P_3=2.08$ is indicated by the green arrow in the pulse stack and the green spectral feature in the 2DFS. Its second harmonic is marked in orange, showing an additional apparent slow negative drifting subpulses.
    }
    \label{fig:Nyquist-simulation}
\end{figure}

A simple fast-rotating carousel is possible to generate pulse stacks which give the illusion of slow drift interrupted by a rapid flicker. To illustrate this, \fig{fig:Nyquist-simulation} shows an simulated pulse stack generated by a carousel with five equal intensity sub-beams and a fixed circulation time $P_4$ of $10.4$.
Therefore $P_3=P_4/n_{\rm sb}=2.08$, corresponding to the prominent spectral feature at $1/P_3\approx0.48~\rm cpp$ as indicated by the green outlined box in the 2DFS in the same figure.
$P_3$ is close to $2$, resulting in chessboard-like pattern in the pulse stack.
The subpulses drift by $1/2.08\approx0.48$ times the spacing between two consecutive sub-beams per pulse period, as illustrated by the green arrow.
Because $P_3$ is not exactly equal to $2$, the fast circulation time creates an appearance that the sub-beams are moving backwards by around 0.04 times the spacing, leading to an apparent slow negative drift, as indicated by the orange arrow.
This resembles a slow drift combined with a rapid flicker.
The slow negative drift can also be identified in the 2DFS as the spectral feature indicated by the orange outlined box.
In fact, this feature ($1/P_3\approx0.04~\rm cpp$) is the alias of the second harmonic of the fundamental drift feature ($1/P_3\approx0.48$).
As a result, not only the $1/P_2$ frequency but also the (unaliased) $1/P_3$ frequency are twice larger than those of the fundamental drift feature. Such a scenario is incompatible with the observed spectral features of PSR~J1514$-$4834 in which the slow drift and flicker frequencies are not harmonically related (see \tab{tab:measure} and \sect{sec:single-periodicity}).

\subsection{Circulation time model}
\label{apdx:sidebands-simulation}

\begin{figure}
 \includegraphics[width=1\linewidth]{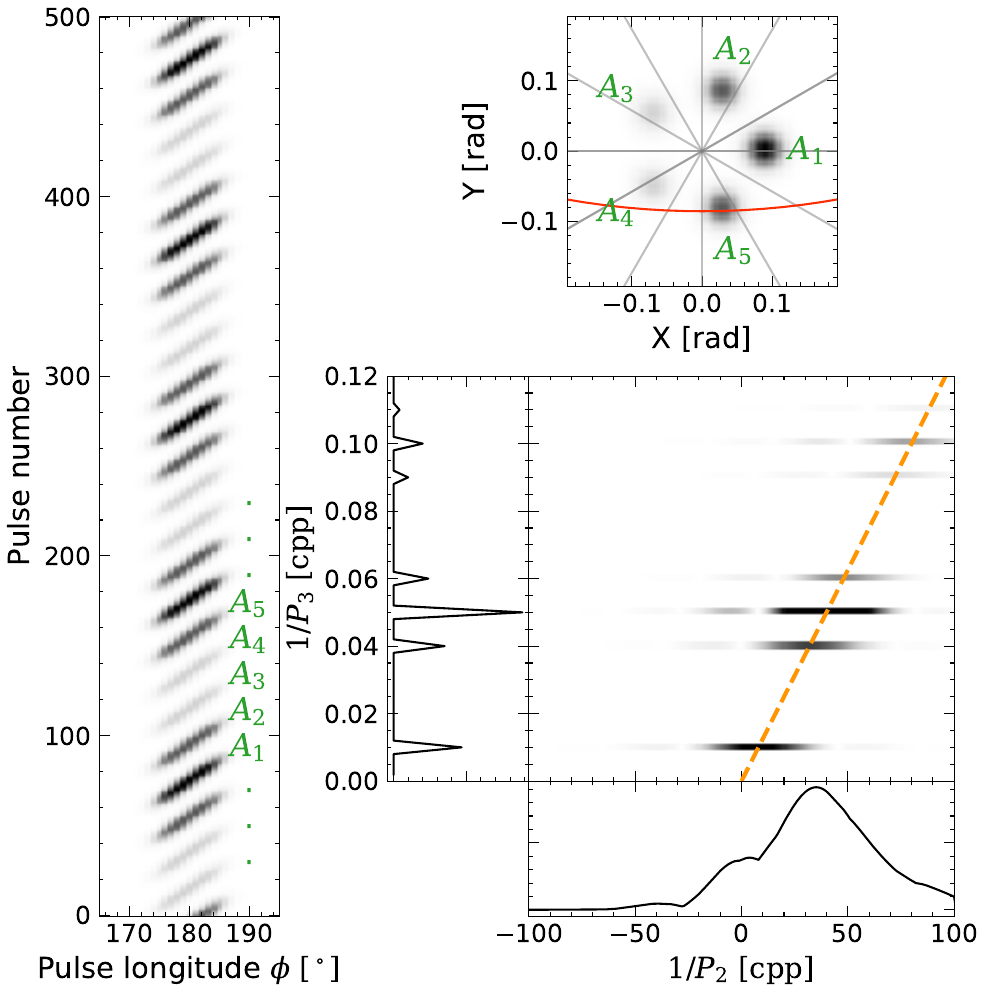}
  \caption{Similar to \fig{fig:Nyquist-simulation}, but here a specific intensity distribution of the sub-beams $A_n$ is adapted to generate a pair of symmetric sidebands ($1/P_3=0.05\pm0.01~\rm cpp$) around the main drift feature ($1/P_3=0.05~\rm cpp$).
  In the 2DFS (bottom-right panel), all the spectral features are aligned with the dashed line passing through the origin, showing that they are harmonically related.}
  \label{fig:sidebands-simulation}
\end{figure}

A carousel with unequal intensity sub-beams is argued to be able to account for the observed sidebands in PSR~B0943$+$10 \citep{dr99, dr01}.
Such a set-up is illustrated in the simulation shown in \fig{fig:sidebands-simulation}. Here we consider a carousel consisting of five sub-beams with different intensities $A_1$, $A_2$, \dots, $A_5$.
As a consequence, the resulting pulse stack has drift bands with unequal intensities.
Indeed, sidebands are observed in the horizontally ($1/P_2$ direction) integrated power side panel of the 2DFS.
A low-frequency ($1/P_3=0.01~\rm cpp$) feature associated with the carousel circulation time appears with a strength comparable to that of the sidebands (at $1/P_3=0.05\pm0.01~\rm cpp$).

The presence of such a circulation feature was also noted in the simulations conducted by \cite{gs03}. 
However, here we highlight that not only the $1/P_3$ frequencies, but also the $1/P_2$ frequencies are harmonically related: all spectral features align with the dashed line in \fig{fig:sidebands-simulation}.
Therefore, the $1/P_2$ frequency of the circulation feature is $n_{\rm sb}$ times smaller than that of the drift feature, in which $n_{\rm sb}$ is the number of sub-beams.
The larger $n_{\rm sb}$ is, the closer the $1/P_2$ frequency of the circulation feature will be to zero.
This can make it appear as a mostly amplitude modulation as inferred in \citep{dr01}.
This conclusion is used in our discussion in \sect{sec:B0943} to disprove such circulation time model for PSR~J1514$-$4834.

%%%%%%%%%%%%%%%%%%%%%%%%%%%%%%%%%%%%%%%%%%%%%%%%%%

% Don't change these lines
\bsp	% typesetting comment
\label{lastpage}
\end{document}